\documentclass[twocolumn,secnumarabic,amssymb, nobibnotes, aps, prd]{revtex4-1}
\usepackage{graphicx}
\usepackage{amsmath}
\usepackage{nccmath}
\usepackage{amssymb}
\usepackage{grffile}
\usepackage{float}
\usepackage{color}
\usepackage{textcomp}
\usepackage[normalem]{ulem}

\begin{document}

\title{Nature of Spinons in 1D Spin Chains}

\author{Teresa Kulka$^1$}

\author{Miłosz Panfil$^1$}

\author{Mona Berciu$^{2,3}$}

\author{Krzysztof Wohlfeld$^{1*}$}

\affiliation{
$^1$Faculty of Physics, University of Warsaw, Ludwika Pasteura 5, 02-093 Warsaw, Poland\\
$^2$Department of Physics and Astronomy, University of British Columbia, 6224 Agricultural Rd, Vancouver, BC V6T 1Z1, Canada\\
$^3$Stewart Blusson Quantum Matter Institute, University of British Columbia, 2355 E Mall, Vancouver, BC V6T 1Z4, Canada
}

\date{\today}%

\begin{abstract}
We provide an intuitive understanding of the collective low-energy spin excitation of the one-dimensional spin-$\frac{1}{2}$ antiferromagnetic Heisenberg chain, known as the spinon. To this end, we demonstrate 
how a single spinon can be excited by adding one extra spin to the ground state.
This procedure accurately reproduces all key features of the spinon's 
dispersion. These follow from the vanishing norm of the excited state which is triggered by the ground state entanglement.
Next, we show that the spinon dispersion can be approximately reproduced if we replace the true ground state with the simplest valence-bond solid. This proves that the spinon of the one-dimensional Heisenberg model can be understood as a single spin flowing through a valence-bond solid.
\end{abstract}

\maketitle

\begin{figure*}[t!]
\centering
\includegraphics[width=0.9\linewidth]{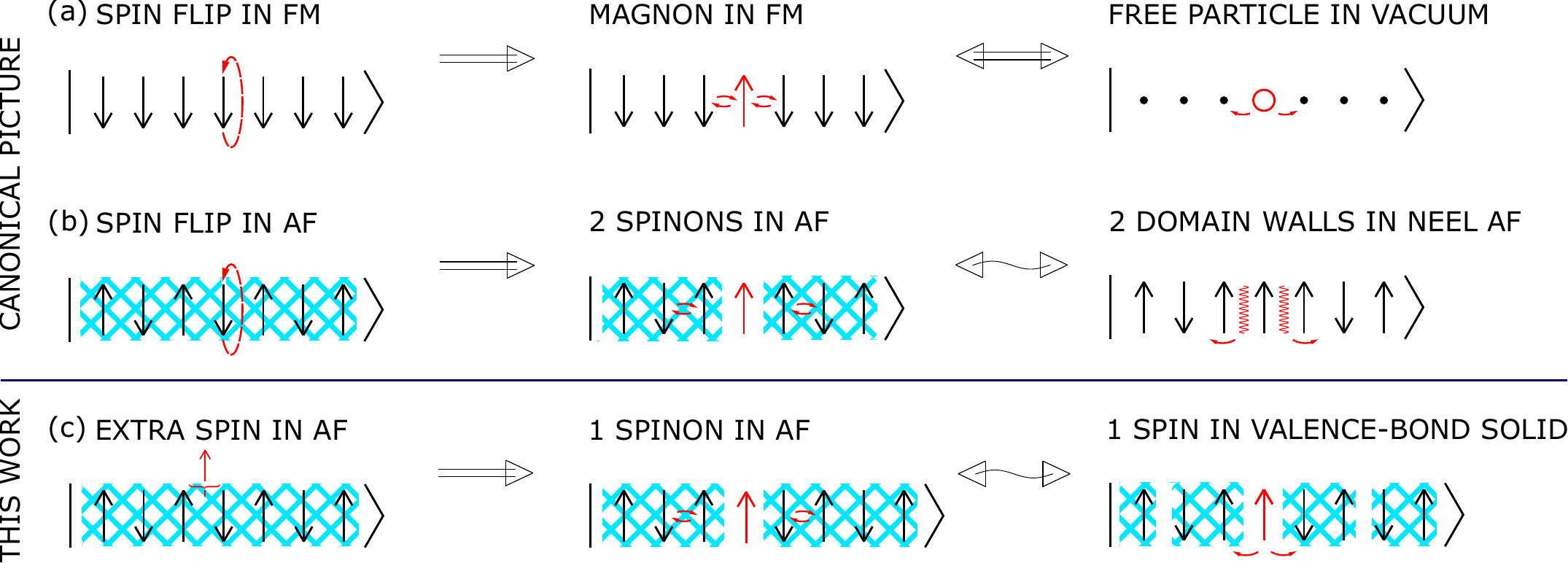}
\caption{
Cartoon views of collective excitations in spin-$\frac{1}{2}$ Heisenberg chains. {\it Panel (a)}: canonical picture arising after flipping one spin in the Heisenberg FM ground state (left subpanel) and triggering a propagating magnon (middle subpanel) which is understood as a free mobile particle in the vacuum (right subpanel).
{\it Panel (b)}: canonical picture arising after flipping one spin in the Heisenberg AF ground state (left subpanel) and triggering two propagating spinons (middle subpanel) which are usually approximately understood as two mobile domain walls in a N\'eel AF (right subpanel).
{\it Panel (c)}: picture introduced in this paper, arising after inserting an additional spin in the Heisenberg AF ground state (left subpanel) and triggering a propagating spinon (middle subpanel) which can be understood as a mobile spin flowing in the valence-bond solid (right subpanel).
Flipped or inserted spins are shown in red; 
entanglement between
two {\it or} more sites is shown by a blue-coloured hatched area.
}
\label{cartoon1}
\end{figure*}

{\it Introduction:}
Some of the most amenable states of interacting quantum matter are those with long-range order~\cite{Khomskii2010}, due to their almost classical, product-like, ground states.
Their low-lying excited states are also rather simple, being best described in terms of 
long-living and weakly interacting 
collective excitations~\cite{Khomskii2010, Venema2016}
whose nature is extremely intuitive. For instance, a magnon in a ferromagnet (FM)~\cite{Auerbach1994}
propagates just like a free particle, see Fig.~\ref{cartoon1}(a).

However, even relatively small
deviations from a ground state with long-range order may lead to surprisingly complex excited states. Probably the best known and most studied  such example is the one-dimensional (1D) spin-$\frac{1}{2}$ antiferromagnetic (AF) Heisenberg model~\cite{1931_Bethe_ZP_71, Auerbach1994}. Its ground state has {\it quasi}-long-range order, with correlations that decay algebraically with distance~\cite{Auerbach1994}. 
Its low-lying excited states are interpreted in terms of spinons, which are rather exotic quasiparticles with a peculiar dispersion that is supported only in  half of the Brillouin zone and with a bandwidth proportional to $\pi$~\cite{1981_Faddeev_PLA_85}, and have so far evaded an intuitive picture.

The complex nature of the spinon is confirmed by the slow historical development of its understanding: while the Bethe Ansatz solution for the Heisenberg model has been known for over 90 years~\cite{1931_Bethe_ZP_71}, it was realised only after 50 years that the spinon carries spin-$\frac{1}{2}$~\cite{1981_Faddeev_PLA_85} instead of the previously assumed spin-$1$~\cite{1962_desCloizeaux_PR_128} of a  magnon-like excitation.
One reason why understanding the spinon's properties was so difficult is  that standard experimental probes do not excite a single spinon.
Instead they excite spin-$1$ flips which decay into the two-spinon (and higher-order) continua~\cite{Tennant1993, Lake2005, Caux2005, Klauser2011, Schlappa2012, Mourigal2013, Ferrari2018}.
This process is usually visualised as the creation of two freely moving domain walls propagating in a N\'eel AF, cf.~Fig.~\ref{cartoon1}(b). As we show here, this simple cartoon is {\em not} a valid picture of the spinon.

In this Letter, we provide a new and intuitive understanding of the spinon's nature. To this end, we first propose a procedure to create a state with a {\it single} spinon excitation by adding one extra spin site to the ground state of a Heisenberg spin chain, cf. Fig.~\ref{cartoon1}(c). Indeed, we show that this accurately reproduces all 
key features of the spinon's dispersion and this excitation carries spin-$\frac12$ by construction.
While our procedure is closely related to the concepts promoted in Refs.~\cite{Sorella1992, Talstra1997, Penc1997, Penc1997b, Matveev2007a, Matveev2007b}, it has an advantage of being simple and and easily amenable to numerical implementation.

We then move to the main results of this Letter. We first
demonstrate that a domain wall propagating through a N\'eel state is 
qualitatively different from a spinon, invalidating the 
popular naive picture of Fig.~\ref{cartoon1}(b).
Remarkably, we find a good approximation of the 
spinon if we replace the true AF ground state with the simplest valence-bond solid. This result is due to 
entanglement, which is a characteristic feature both of the Heisenberg spin chain ground state~\cite{PhysRevLett.90.227902,Sato_2006,XXX_density_matrix,10.21468/SciPostPhys.8.3.046,PhysRevA.102.042206} {\it and} of the valence-bond solid~\cite{Alet2007}
and which is absent in the  N\'eel state.
Altogether, this enables us to draw the following intuitive explanation of the key spinon properties:
(i) the limited support of the spinon dispersion stems from entangled fluctuating spins 
leading to destructive interference of the spinon wave for $|q|>\pi/2$;
(ii) 
the low-energy linear dispersion arises {\it both} 
because of (i) {\it and}
because a single spin can hop between 
adjacent sites in the valence-bond state, cf. Fig.~\ref{cartoon1}(c).

\emph{The model:}
The Hamiltonian of the Heisenberg spin chain is~\cite{1931_Bethe_ZP_71,korepin_bogoliubov_izergin_1993}
\begin{equation} \label{Heisenberg}
	H_L = J \sum_{j=1}^L \vec{\sigma}_j \cdot \vec{\sigma}_{j+1},
\end{equation}
where $\vec{\sigma}_j = (\sigma_j^{x}, \sigma_j^{y}, \sigma_j^{z})$ are spin-$\frac{1}{2}$ Pauli matrices acting on the $j$-th site of the chain. We assume periodic boundary conditions, $\vec{\sigma}_{L+1} = \vec{\sigma}_1$. In the antiferromagnetic case $J>0$ and in the following we set $J=1$. 

The 1D Heisenberg model is exactly solvable with the Bethe Ansatz~\cite{korepin_bogoliubov_izergin_1993}. For the AF case and even $L$, it has a unique ground state and its magnetization $M = \sum_{j=1}^L \sigma_j^z$ is zero. For odd $L$, the ground state is doubly degenerate. The low-energy excitations above the ground state are obtained by exciting an even number of spinons. A single spinon has spin-$\frac{1}{2}$ and its dispersion relation
\begin{equation} \label{spinon_dispersion}
	\varepsilon(q) = \frac{\pi}{2} \cos q, \quad 
 q \in \bigg[-\frac{\pi}{2}, \frac{\pi}{2}\bigg] 
\end{equation}
 is supported only in half of the Brillouin zone~\cite{1981_Faddeev_PLA_85}. While the Bethe Ansatz  provides an exact description of the spinon, it lacks a physically intuitive picture.

\emph{How to create a single spinon:}
We now present a general method of constructing elementary excitations for any spin chain. As we show next, when applied to the Heisenberg chain, its result is a single spinon excitation.

\begin{figure*}[t!]
\centering
\includegraphics[width=0.9\linewidth]{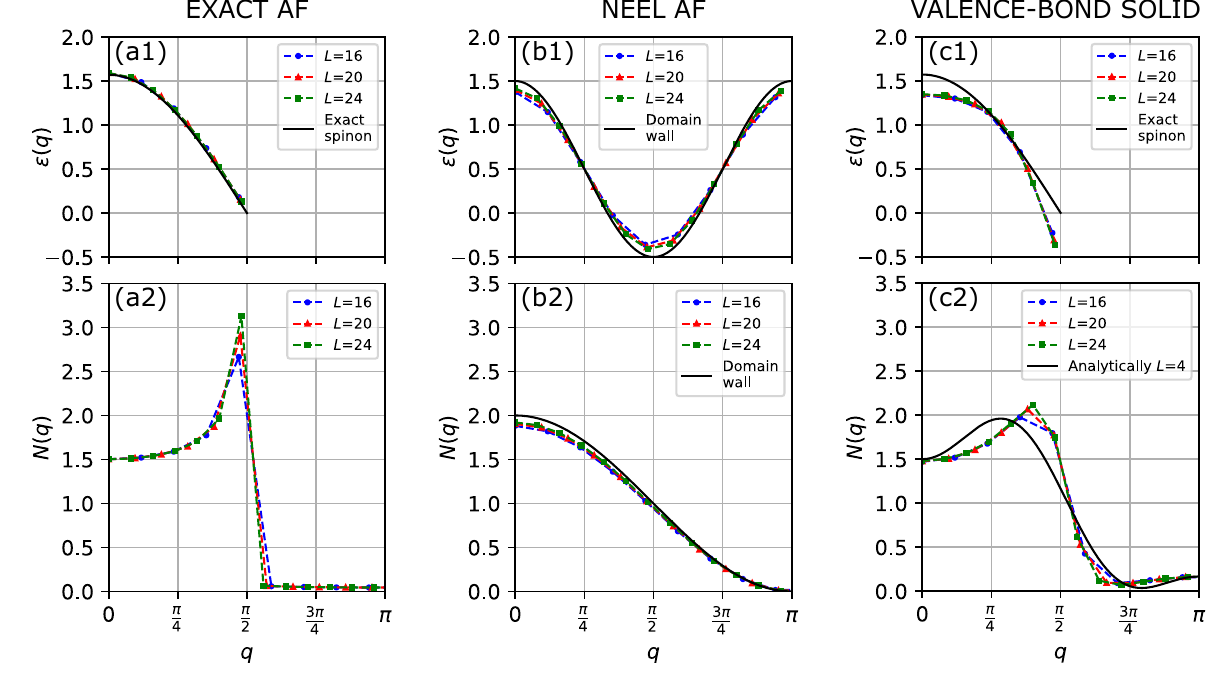}
\caption{
Top panels show the dispersion $\varepsilon (q)$ of the excitation, defined in Eq.~(\ref{dispersion}), while the bottom panels show the norm $N(q)=\langle \psi(q)|\psi(q) \rangle$. The excited state $|\psi(q) \rangle$ of Eq.~(\ref{psiq}) is built: from the {\it exact AF} ground state $|\Psi_0 \rangle$ obtained by numerically diagonalising the Heisenberg Hamiltonian on $L$ sites [panels (a1)-(a2)]; by using the {\it N\'eel AF} state of Eq.~\eqref{neel}  instead of $|\Psi_0 \rangle$ [panels (b1)-(b2)]; by using  the spin {\it valence-bond solid} state of Eq.~\eqref{DW} instead of $|\Psi_0 \rangle$ [panels (c1)-(c2)]. 
Results for chains of various length $L$ are shown by symbols (the connecting dashed lines are guides to the eye). The thick black solid lines show the relevant analytical expressions, for comparison purposes (see text for more details).
}
\label{results}
\end{figure*}

For the spin chain of length $L$, we assume that we know the expansion of its ground state in the local spin basis:
\begin{equation}\label{psi0}
|\Psi_{0}\rangle=\sum_{j}A_{j}|j\rangle.
\end{equation}
In principle $j$ extends over $2^L$ possible local states, however in practice various selection rules force many $A_j$ to vanish. For example, for $L$ even and if $|\Psi_0\rangle$ 
has total magnetization $M=0$, the sum includes maximally $\binom{L}{L/2} \sim \frac{2^L}{\sqrt{L}}$ states.
The coefficients $A_j$ can be obtained from exact diagonalization or, for integrable spin chains, by the coordinate Bethe Ansatz~\cite{korepin_bogoliubov_izergin_1993}.

Next, we insert a new site with a spin {\it up} between sites $m-1$ and $m$ of the original chain, thus extending the chain length to $L+1$ sites, cf. left panel of Fig.~\ref{cartoon1}(c).
Specifically, for each state $|j\rangle$ expressed in the local spin basis this creates a uniquely defined state $|j(m)\rangle$, $m=0, \dots, L$,
\begin{equation}
	|j\rangle = |\! \underbrace{\downarrow \downarrow \cdots  \uparrow}_{L} \rangle \;\rightarrow\; |j(m) \rangle = |\! \underbrace{\downarrow \downarrow \cdots}_{m-1}  \uparrow \!\!\!\underbrace{\cdots \uparrow}_{L-m+1} \!\!\rangle.
\end{equation}
The resulting basis $|j(m)\rangle$ is over-complete because it contains $2^L (L+1)$ states.
We use it to define an approximate excited state
\begin{equation}
|\Psi(m)\rangle = \sum_{j}A_{j}|j(m)\rangle.
\end{equation}
An important feature of the construction, further discussed below, is that $|\Psi(m)\rangle$ is not normalized. Denoting by $|\bar{\Psi}(m)\rangle$ its normalized counterpart, we find the first two moments of the local spin operators to be
$\langle\bar{\Psi}(m)|\sigma_j^z|\bar{\Psi}(m)\rangle=\frac{\delta_{jm}}{2}$, $\langle \bar{\Psi}(m)|(\sigma_m^z)^2 |\bar{\Psi}(m)\rangle  = \frac{1}{4}$, and for $j \neq m$: $\langle \bar{\Psi}(m)|(\sigma_j^z)^2 |\bar{\Psi}(m)\rangle = \langle \Psi_0|(\sigma_1^z)^2 |\Psi_0\rangle $, where in the last equality we used the translational invariance of the ground state.  This result can be visualized as a spin {\it up} frozen at the $m^{th}$ site of the $L+1$-site chain, in the sea of fluctuating spins on all other sites. These fluctuations are identical to the ground state fluctuations of the $L$-site chain, by  construction. This picture is in accordance with the concept of elementary excitations which are created from the ground state, and inherit rather than rebuild the correlations present in it.

To define a dispersion relation we introduce a Fourier transform with momentum $q=\frac{2\pi n}{L+1}$, $ -L/2 \leqslant n \leqslant L/2$, $n\in\mathbb{Z}$:
\begin{equation} \label{psiq}
|\Psi(q)\rangle=\frac{1}{\sqrt{L+1}}\sum^{L}_{m=0}e^{iqm}|\Psi(m)\rangle.
\end{equation}
The state $|\Psi(q)\rangle$ is not an eigenstate of the Hamiltonian, so we define its energy as the expectation value and calculate it with respect to the ground state energy $E_0^{L+1}$ of a system of size $L+1$:
\begin{equation} \label{dispersion}
	\varepsilon(q)=\frac{\langle\Psi(q)|H_{L+1}|\Psi(q)\rangle}{\langle\Psi(q)|\Psi(q)\rangle} - E^{L+1}_{0},
\end{equation}
where $H_{L+1}$ is the Hamiltonian for a system of size $L+1$.

\emph{Spinon in the AF Heisenberg chain:} We now show that if we apply the procedure outlined above to the Heisenberg spin-$\frac12$ chain, we recover the spinon dispersion relation~\eqref{spinon_dispersion} including the correct support of half of the Brillouin zone. 
The amplitudes $A_j$ are computed from exact diagonalization. The results are shown in Fig.~\ref{results}(a1). We observe that the spinon dispersion relation $\varepsilon(q)$, shown by the black solid line, is recovered with excellent precision. To understand this result, we also plot separately the denominator of Eq.~\eqref{dispersion}, namely:
\begin{equation}\label{NE}
 N(q)=\langle\Psi(q)|\Psi(q)\rangle. 
\end{equation}
We find that  $N(q)$  
vanishes 
(similarly to \cite{Talstra1997}, its value is
negligible)
for $|q| > \pi /2 $, cf.~Fig.~\ref{results}(a2). States of zero norm are not physical and therefore we exclude them from further considerations. The consequence is that the spinon 
only exists in  half of the Brillouin zone. 

In our procedure, the correlations present in the spinon state are 
inherited from the ground state. This leads to the natural question as to which correlations, of all those present in the exact ground state, are most relevant to the spinon. To answer this question we consider two 
approximations to the ground state: (i) the  lowest  energy classical configuration, {\it i.e.} the N\'eel state, and (ii) the simplest valence-bond solid.

\emph{(i) Domain wall approximation:} In the first approximation,  we therefore replace:
\begin{align}\label{neel}
|\Psi_{0}\rangle \approx |\!\uparrow\downarrow\uparrow\downarrow\uparrow\downarrow\uparrow\downarrow\ldots\rangle.
\end{align}
We then apply the procedure to this wavefunction, see left subpanel of Fig.~\ref{cartoon1}. 
The results for the corresponding $\varepsilon(q)$ and $N(q)$ 
are shown in Fig.~\ref{results}(b1-b2).

This dispersion $\varepsilon(q)$ reproduces the domain wall dispersion found in the Ising-like AF~\cite{Villain1975},
$\varepsilon_{dw}(q)=\frac{1}{2}+\cos(2q)$,
shown as a solid black line in Fig.~\ref{results}(b1),
instead of the spinon dispersion. This difference is directly caused by the use of the N\'eel state instead of the true ground state in the procedure.

\emph{(ii)
Additional spin
in valence-bond solid:} As a second approximation, we replace the ground state with a valence-bond solid:
\begin{equation}\label{DW}
	|\Psi_0\rangle\approx\frac{|\!\uparrow\downarrow\rangle-|\!\downarrow\uparrow\rangle}{\sqrt{2}}\otimes\frac{|\!\uparrow\downarrow\rangle-|\!\downarrow\uparrow\rangle}{\sqrt{2}}\otimes\ldots,
\end{equation}
leading to a spin being added (in our procedure) to a state consisting of singlets on neighbouring sites,
cf. right subpanel of ~Fig~\ref{cartoon1}(c).

The corresponding $\varepsilon(q)$, $N(q)$ 
are shown in Fig.~\ref{results}(c1-c2). The dispersion relation is qualitatively similar to the spinon dispersion (black solid line).
Quantitatively, the value at $q=0$ is underestimated and the value at $q=\frac{\pi}{2}$ is negative, reflecting the fact that the valence-bond state is an approximation.
Indeed, its average energy per site is $-0.3750$, significantly larger than the exact ground state energy per site of $-0.4438$~\cite{Auerbach1994}.

We emphasize that this approximation 
recovers the linear low-energy dispersion $\varepsilon(q) \propto |{\pi\over 2} - q|$ as $q \rightarrow {\pi \over 2}$ and
the extended region  $|q| \gtrsim \pi/ 2$ where the norm $N(q)$ nearly vanishes. 
The latter is
similar to the result of exact diagonalization presented in Fig.~\ref{results}(a2).
Interestingly, it is qualitatively the same
as an exact result for a short $L=4$ chain, see Fig.~2(c2) and Sec. I of \cite{SM}, which further supports the close connection between the valence-bond approximation and the exact result.

Contrasting the two approximations reveals the crucial role played by 
a particular form of the ground state correlations --
that is quantum entanglement --
in determining the spinon properties. We further discuss the two main characteristics below.

\emph{The limited support of the spinon dispersion:}  By excluding from the Hilbert space states of zero norm, the question of the support of the spinon dispersion is translated into the question of a (nearly) vanishing $ N(q)$. To understand the key role played by the ground state entanglement, let us consider the structure of $|\Psi(q)\rangle = \sum_{j'} B_{j'}(q) |j'\rangle$ expressed in the spin basis of the longer chain. Any new spin configuration $|j'\rangle$ is obtained from all of the original states of the shorter chain $|j\rangle$ that differ from $|j'\rangle$ only through a missing spin {\it up}. Insertion of the missing spin {\it up} will generate $|j'\rangle$, however because these inserted spins {\it up} are \emph{at different positions}, different contributions are multiplied by different Fourier factors $e^{i q m}$ in $B_{j'}(q)$. 

Crucially, a configuration $|j'\rangle$ may
originate from multiple $|j \rangle$'s in multiple ways. This gives rise to the momentum dependence of the norm $N(q)$, even for a product ground state. For example, consider a N\'eel ground state, i.e. a unique $|j\rangle= |\!\! \uparrow \downarrow \uparrow \downarrow \dots \rangle$, and let $|j'\rangle = |\!\!\uparrow \uparrow \downarrow \uparrow \downarrow \dots\rangle$. This state can be obtained from $|j\rangle$ by inserting the extra spin {\it up} either before or after the first spin {\it up} already present. The Fourier coefficients of these two contributions differ by $e^{iq}$, $B_{j'}(q) = 1 + e^{iq}$, therefore for the N\'eel state
the norm vanishes at $q = \pi$, as seen in Fig.~\ref{results}(b2).

In an entangled ground state, however, a
$|j' \rangle$ can be reached by adding a spin 
in different places in different $|j \rangle$
configurations. 
To illustrate this, consider the  $L=4$-site
Heisenberg chain, see Sec. I of \cite{SM} for details. Its ground state is entangled and includes, {\it inter alia}, the two contributions: (i) $| \uparrow \downarrow \uparrow \downarrow \rangle $,  and (ii)
$- \frac12 | \uparrow \uparrow \downarrow \downarrow \rangle$. 
If we add a spin {\it up}, before or after the first spin, in (i)
and after the third spin in (ii), respectively, we reach the
same $|j'\rangle = |\!\!\uparrow \uparrow \downarrow \uparrow \downarrow \rangle$ with different phase factors, leading to $B_{j'}(q) = 1 +e^{iq}- e^{3iq} / 2$. 
 
It is crucial in the above-obtained equation that the different factors have different phases.
On one hand, it leads to the nearly vanishing norm $N(q)$ in an extensive region of momenta $q$, see Sec. I of [SM] for more details. On the other hand, such a non-trivial relative phase factor in front of the Fourier coefficient $e^{imq}$ can only be obtained for the ground state that is entangled. 
Physically this means that the `fluctuating' spins stemming from the entangled ground state,
lead to a (nearly) destructive interference in the spinon wave and thus to the (nearly) vanishing of its norm $N(q)$ for some momenta.
Last but not least, we put this qualitative analysis on firm ground by proving that $N(0)$ is an entanglement witness: As shown in Sec. II of \cite{SM}, for a translationally invariant superposition of product states of zero magnetization $N(0) \geq 2$. 
Therefore, $N(0) < 2$ signals an entangled ground state.

However, there is more to the norm $N(q)$ than that, as it can behave as a function of $q$ in rather extreme ways:
it may vanish for an extensive range of momenta $q$, or may diverge. This happens for the exact ground state of the Heisenberg model in the thermodynamic limit, though not for the finite Heisenberg chains or the valence-bond solid [cf. panels (a) and (c) of Fig.~\ref{results}]. This observation is in agreement with the notion of the short-range entanglement present in the ground state of the Heisenberg model in the thermodynamic limit~\cite{Vidal2003} and the semi-local (`minimal') entanglement of the valence-bond ground state~\cite{Alet2007}. This shows that $N(q)$ contains rich information about the structure of the state and further studies are required to fully harvest its potential.

\emph{Dispersion linear in momentum:}
The other notable feature of the spinon dispersion is its linear dependence on  $q$ in the low-energy limit. 
It can be understood as a result of (i) the vanishing 
compact support  for $|q| > \pi /2$  that is due to ground state entanglement (see above), {\it and} (ii) the $\propto \cos q$ dispersion relation arising from nearest
neighbor hopping of the spinon, discussed next.

To this end we note that the linear dispersion relation {\it is (is not)} reproduced when the single spin is added to the {\it valence-bond (N\'eel)} state, cf. Fig.~\ref{results}(b-c).
This is because in the valence-bond state the 
singlets  `resonate' with each other as well as with the added spin and move by one site with a single action of the Hamiltonian.
On the other hand, a domain wall in the Neel state can only move between next nearest neighbour sites.

\emph{Outlook:} 
Following the  publicly available draft of this paper, the procedure to excite single spinons was just tested experimentally using STM on a
fabricated nanographene Heisenberg  spin chain~\cite{Zhao2025}.
In particular, the Authors of ~\cite{Zhao2025} managed to observe a single spinon standing wave in an STM experiment. This not only confirms the findings of this paper but also shows that it can indeed be experimentally implemented. Finally, we suggest that also particular pump-probe experimental techniques may be able to implement this procedure, for more details see~Sec.~IV of~\cite{SM}.

The postulated procedure is universal and can 
be applied to any 1D spin model,
cf. Sec. III of \cite{SM} for a short overview.
In particular, it can be used to study excitations of the 1D FM chain, see Fig. 2(a-b) of \cite{SM}: in this case a single magnon is excited in the procedure. This further shows that the spinon is {\it not} exotic because it can be excited in quite a special manner {\it but} it is the entanglement nature of the antiferromagnetic ground state that is crucial here.
We suggest that in the future, this procedure can be used to explore the nature of the elementary excitations in frustrated spin systems, such as the recently discussed problem of marginally irrelevant interactions between spinons in $J_1-J_2$ spin chains~\cite{Keselman2020}.
Last but not least, our understanding is that the above-mentioned experimental techniques can in principle be carried out also in higher dimensions. This way the nature of single-spinon excitations beyond 1D can be revealed.

{\it Conclusions:} Our work confirms a paradigm that is rarely explicitly formulated, according to which a collective excitation generally inherits its properties from the ground state. This paradigm is probably best exemplified by the Goldstone theorem and the close relation between the broken-symmetry long-range ordered ground state and the gapless collective excitations~\cite{Beekman2019}.

Here we used this paradigm to understand a system that does not have spontaneous symmetry breaking and where the Goldstone theorem is not valid, hence the connection between the ground and the excited states is less clear. Our successful explanation of the properties of the spinon excitation shows that such a connection is possible in complex correlated systems as well. The resulting features of the elementary excitations are then controlled by the entanglement in the ground state.

\emph{Acknowledgements:}
Authors acknowledge the support of National Science Centre in Poland under the projects
2016/22/E/ST3/00560 (K.W.), 2024/55/B/ST3/03144 (K.W), 2018/31/B/ST3/03758 (T.K.), 2018/31/D/ST3/03588 (M.P.), 2022/47/B/ST2/03334 (M.P.). 
K.W. acknowledges support from the Excellence Initiative of the University of Warsaw (the `mikrogrant' programme).
M.B. acknowledges support from the Canada First Research Excellence Fund and the 
Natural Sciences and Engineering Research Council of Canada.
We thank Federico Becca, Akira Furusaki and Karlo Penc for stimulating discussions and all the organisers of the KITP program ``A New Spin on Quantum Mangets'' for allowing us to present this work during the program. We thank the Referee for bringing Ref.~\cite{Talstra1997} to our attention, Karlo Penc for pointing out references ~\cite{Sorella1992, Penc1997, Penc1997b}, and Akira Furusaki for pointing out references ~\cite{Matveev2007a, Matveev2007b}. This research was supported in part by grant
NSF PHY-2309135 to the Kavli Institute for Theoretical
Physics (KITP).
This research was carried out with the support of the Interdisciplinary Centre  for Mathematical and Computational Modelling at University of Warsaw (ICM UW) under grant no. G73-29.


\onecolumngrid

\setcounter{equation}{0}
\setcounter{figure}{0}
\renewcommand{\thetable}{S\arabic{table}}
\renewcommand{\theequation}{S\thesection.\arabic{equation}}
\renewcommand{\thefigure}{S\arabic{figure}}
\setcounter{secnumdepth}{2}

\newpage

\begin{center}
{\Large Supplemental Material \\ 
Nature of spinons in the 1D spin chains}
\end{center}
\tableofcontents

\section{Exact result for the $4$-site Heisenberg chain: relation between the norm $N(q)$ and entanglement}

We discuss below the AF Heisenberg chain on $L=4$ sites. Here the {\it resonating} valence-bond state is an exact ground state of the Heisenberg model. Up to normalization,
\begin{equation}
    |\Psi_0\rangle = \frac{|\!\uparrow\downarrow\rangle-|\!\downarrow\uparrow\rangle}{\sqrt{2}}\otimes\frac{|\!\uparrow\downarrow\rangle-|\!\downarrow\uparrow\rangle}{\sqrt{2}} + {\rm translation} .
\end{equation}
By following our procedure, we find:
\begin{equation}
    |\Psi(q)\rangle = \frac{1+e^{iq} - \frac{e^{3iq}}{2}}{\sqrt{3}} |A\rangle - \frac{1 + e^{iq} + e^{2iq}}{2\sqrt{3}} |B\rangle
\end{equation}
with $|A\rangle = |\!\!\uparrow\uparrow\downarrow\uparrow\downarrow\rangle + ({\rm translations})$ and $|B\rangle = |\!\!\uparrow\uparrow\uparrow\downarrow\downarrow \rangle + ({\rm translations})$. All the states in $|A\rangle$ and $|B\rangle$ are orthonormal to each other. The resulting norm:
\begin{equation} \label{norm_N_4}
  N(q)= 1 + \cos q - \frac{1}{6} \cos(2q) - \frac{1}{3} \cos(3q)
\end{equation}
shows, {\it inter alia}, a considerable suppression of $N(q)$ in an extended interval from around $\frac{\pi}{2}$ to $\pi$, see black solid line in
Fig.~2(c2) of the main text and discussion below. Of course, for $L=4$ this result is only meaningful at multiples of $q=\frac{2\pi}{5}$, but similar arguments hold for longer chains within the valence-bond approximation.

Having obtained an exact expression for $N(q)$, \eqref{norm_N_4}, we are now ready to investigate the origin of its two main striking properties: (i) a relative enhancement  of the $4$-site Heisenberg chain norm, with a pronounced maximum for the small momenta $q \lesssim \pi/2 $,  w.r.t. the norm obtained once a spin is added to the Neel antiferromagnet, that is w.r.t. $N(q) = 1+\cos(q)$ (see main text); (ii) a relative suppression  of the $4$-site Heisenberg chain norm, with a pronounced minimum for the small momenta $q \gtrsim \pi/2 $,  w.r.t. the norm obtained once a spin is added to the Neel antiferromagnet, that is w.r.t. $N(q) = 1+\cos(q)$ (see main text).

To this end, let us first note that a sum rule $ \sum_{ 0 <q <\pi} N(q) = \pi$ means that once we e.g. understand the relative enhancement 
for some momenta, then the relative suppression for the other momenta follows. Therefore, we concentrate on investigating the former effect.

A quick inspection of Eq.~\eqref{norm_N_4} shows that the term responsible for obtaining such a relative enhancement in $N(q)$ for small momenta (and, conversely, also a relative suppression for large momenta) is the term $ \propto - \cos (3q) $. A closer look at the procedure of adding an extra spin shows that, in order to obtain such a contribution to the norm, there must exist two product states in the ground state which contribute to $ |\Psi_0 \rangle $ with opposite sign and that they must be identical to each other after adding an extra spin in our procedure, modulo a phase factor $e^{3 i q}$. 
This is fulfilled for instance by such two contributions to the exact ground state
$| \Psi_0 \rangle$ that, upon adding an extra spin site at particular location, lead to the same states which differ by $e^{3iq}$ in our procedure:
\begin{align} \label{eq:updown}
   | \uparrow \downarrow \uparrow \downarrow \rangle \quad &\rightarrow  \quad  |  \tilde{\uparrow} \uparrow \downarrow \uparrow \downarrow \rangle,   \\     - \frac12 | \uparrow \uparrow \downarrow \downarrow \rangle \quad    &\rightarrow \quad    - \frac12 e^{3 i q} |  \uparrow \uparrow \downarrow \tilde{\uparrow} \downarrow \rangle,\label{eq:upup}
\end{align}
where the tilde sign denotes the added spin. Note that when the spin up is added {\it after} the first spin in
$  | \uparrow \downarrow \uparrow \downarrow \rangle $, then we obtain the same configuration 
$ |  \uparrow \uparrow \downarrow {\uparrow} \downarrow \rangle $---albeit with a factor of $e^{i q}$.  

Identifying the type of product states in $|\Psi_0 \rangle$ that lead to the relative enhancement of the norm $N(q)$ for small momenta, allows us to intuitively understand the origin of this phenomenon: namely, it follows from the valence bond nature of the ground state---the two `fluctuating singlets' allow for the onset of ferromagnetic 
domains with opposite magnetisation {\it and} a (relative) negative phase contribution to the ground state, see Eq.~\eqref{eq:upup}. Note that the relative negative phase is just as crucial as the existence of ferromagnetic domains and that such a phase difference can only be realised once $|\Psi_0 \rangle$ is {\it not} a product state. Altogether, this shows how specific properties of the norm $N(q)$ of the state constructed by adding one extra spin can be related to the entangled nature of the spin chain ground state $| \Psi_0 \rangle$.

\section{The norm $N(0)$ as an entanglement witness}

In this section we demonstrate a relation between $N(0)$ and the entanglement present in the ground state $|\Psi \rangle$. More specifically, we derive an exact lower bound $N(0) \geq 2$ for product states. This result further promotes $N(q)$ to be a quantity providing useful information about the system and not merely a normalization. To this end, we reformulate the construction of the excited states by introducing operatorial formalism.

We start by introducing operators $c_j$ that add a site between sites $j-1$ and $j$ and occupy it with a spin up. They act as the identity on the remaining lattice sites. Similar operators were considered before in the context of lattice supersymmetry~\cite{Hagendorf_2017}. Using the notation introduced in the main text,
\begin{equation}
    |\Psi(m)\rangle = c_m |\Psi_0\rangle.
\end{equation}
From the local operators $c_j$ we built operators $C_L$ and $\bar{C}_L$ which act on the whole chain
\begin{equation}
    C_L = \sum_{j=0}^{L-1} c_j, \qquad \bar{C}_L = \sum_{j=0}^L c_j.
\end{equation}
The result of their action on a single state is a superposition of states with an extra spin-up inserted before each lattice site (for $C_L$) and additionally also after the last site (for $\bar{C}_L$). With this formalism the excited state at $q=0$ can be written as
\begin{equation}
    |\Psi(q=0) \rangle = \frac{1}{\sqrt{L+1}} \bar{C}_L |\Psi_0\rangle.
\end{equation}
and 
\begin{equation}
    N(0) = \frac{1}{L+1} \frac{\langle \Psi_0| \bar{C}_L^{\dagger} \bar{C}_L |\Psi_0\rangle}{\langle \Psi_0|\Psi_0\rangle},
\end{equation}
where we explicitly included the norm. 

In the main text we consider systems with periodic boundary conditions. Their ground states are translationally invariant which has important consequences for the evaluation of $N(0)$. Any translationally invariant state can be generated from a {\em seed state} by considering a superposition of all its translations. To formalize this, let us define operator $t_n$ translating a state by $n$ sites to the right and respecting the periodic boundary condition. Similarly, we define $T_L = \sum_{n=0}^{L-1} t_n$. For any translationally invariant state $|\Psi_0\rangle$ there exists a state $|\Psi\rangle$ such that $|\Psi_0\rangle = T_L |\Psi\rangle$ where $|\Psi\rangle$ plays a role of a seed state. Note that if two states $|\Psi_1\rangle$ and $|\Psi_2\rangle$ are related by a translation, i.e. there exist $m = 1, \dots L-1$ such that $t_m|\Psi_1\rangle = |\Psi_2\rangle$ then $T_L |\Psi_1\rangle = T_L |\Psi_2\rangle$. This implies that a translationally invariant state can be generated from any of its seed states.

Being seed states of the same state $|\Psi_0\rangle$ is an equivalence relation and therefore any translationally invariant state can be associated with an equivalence class of its seed states that we denote $[|\Psi_0\rangle]$ and which is characterized by the following defining property
\begin{equation} \label{translation_equivalance}
    T_L|\Psi_1\rangle = T_L|\Psi_2\rangle, \qquad |\Psi_i\rangle \in [|\Psi_0\rangle].
\end{equation}
We will use now this notion to reformulate the computation of $N(0)$ for a translationally invariant states. To avoid confusions we denote by $\mathcal{N}$ the value of $N(0)$ for translationally invariant state. 

As discussed above, for a translationally invariant state $|\Psi_0\rangle$ there exist $|\Psi\rangle$ such that $|\Psi_0\rangle = T|\Psi\rangle$. Therefore
\begin{equation} \label{N_first}
    \mathcal{N}_{\Psi_0} = \frac{1}{L+1} \frac{\langle \Psi| T_L^{\dagger} \bar{C}_L^{\dagger} \bar{C}_L T_L|\Psi\rangle}{\langle \Psi| T_L^{\dagger} T_L |\Psi\rangle}.
\end{equation}
From the property~\eqref{translation_equivalance} it follows that $\mathcal{N}$ is the same for any $|\Psi\rangle$ in the equivalance class $[|\Psi_0\rangle]$ of $|\Psi_0\rangle$. This implies that to compute $\mathcal{N}$ for a translationally invariant state we can identify one of its seeds and then use equation~\eqref{N_first}. This equation still explicitly symmetrizes over all translations but we will now rewrite in a form more suitable for further computations. In this process the above mentioned symmetry becomes less obvious.

We start by describing some properties of operators $C_L$, $\bar{C_L}$ and $T_L$. 
First we observe that $T_L$ is hermitian and has a simple property $T_L^2 = L T_L$. Furthermore the following relations $\bar{C}_L T_L = T_{L+1} C_L$ and $\bar{C}_L T_L = T_{L+1} c_m  T_L$, with $m=0, \dots, L$, hold. Specifically, the first relation shows why in our procedure we insert the extra spin before \emph{and} after the chain. This is because when this operation is applied to a translationally invariant state it produces a translationally invariant state. 

Using the properties of operators $C_L$ and $T_L$ we can prove a simple property of $\langle \Psi_0| \bar{C}_L^{\dagger} \bar{C}_L | \Psi_0 \rangle$ that, due to the translational invariance, position of one of the insertions can be fixed,
\begin{align}
    \langle \Psi_0| \bar{C}_L^{\dagger} \bar{C}_L | \Psi_0 \rangle = \langle \Psi| T_L \bar{C}_L^{\dagger} \bar{C}_L T_L | \Psi \rangle = \langle \Psi| C_L^{\dagger} T_{L+1} T_{L+1} c_m T_L | \Psi \rangle = (L+1)\langle \Psi| \bar{C}_L^{\dagger} T_{L+1} c_m T_L | \Psi \rangle
\end{align}
This leads to an equivalent expression for $\mathcal{N}$
\begin{equation}
    \mathcal{N} = \frac{1}{L} \frac{\langle \Psi| T_L \bar{C}_L^{\dagger} c_0 T_L | \Psi \rangle}{\langle \Psi|T_L| \Psi \rangle}.
\end{equation}
We will now use this representation to prove a bound on $\mathcal{N}$ if $|\Psi\rangle$ is a product state and has an inner symmetry under translations. To this end we introduce a notion of states partially invariant under translations. That is for such state $|\Psi\rangle$ there exists a minimal non-zero $m$ such that
\begin{equation} \label{partial_invariance}
    t_m |\Psi \rangle = |\Psi\rangle. 
\end{equation}
We observe that the property of the partial invariance is shared by all the states in the same equivalance class. 
The consequence of this is that if additionally $|\Psi\rangle$ is a product state then it can be written as a tensor product of its elementary cells 
\begin{equation}
    |\Psi\rangle = \bigotimes_{i=1}^k |\psi\rangle,
\end{equation}
with $L = m k$. 

{\bf Bound for partially invariant product states:} 
We will now show that 
\begin{equation} \label{app_bound_cell}
    \mathcal{N}_{\Psi_0} \geq N_{\psi_0}.
\end{equation}

Assuming~\eqref{partial_invariance} we have $T_L |\Psi\rangle = k T_m |\Psi\rangle$. We then find
\begin{align}
    \mathcal{N}_{\Psi_0} &= \frac{1}{L} \frac{\langle \Psi | T_L \bar{C}_L^{\dagger} c_0 T_L |\Psi \rangle}{\langle \Psi | T_L |\Psi \rangle} = \frac{1}{m}\frac{\langle \Psi | T_m \bar{C}_L^{\dagger} c_0 T_m |\Psi \rangle}{\langle \Psi | \Psi \rangle} \nonumber \\
    &\geq \frac{1}{m}\frac{\langle \Psi | T_m \bar{C}_m^{\dagger} c_0 T_m |\Psi \rangle}{\langle \Psi | \Psi \rangle} = \frac{1}{m}\frac{\langle \psi | T_m \bar{C}_m^{\dagger} c_0 T_m |\psi \rangle}{\langle \psi | \psi \rangle} = \mathcal{N}_{\psi_0},
\end{align}
where the inequality is a consequence of truncating the action of $C_L^{\dagger}$ to the first $m$ lattice sites. This reduces $C_L^{\dagger}$ to~$C_m^{\dagger}$. 

This proves that $\mathcal{N}$ for a seed $|\Psi\rangle$ which is a product state can be bounded by $\mathcal{N}$ of the elementary cell $|\psi\rangle$ of the seed. This is important because if we now want to find further bounds on $\mathcal{N}$ we can focus on states which do not have any invariance under translations which simplifies the analysis.
To this end, we observe that for state $|\Psi_0\rangle$ such that its seed $|\Psi\rangle$ does not have any translational symmetry the formula~\eqref{N_first} can be reduced to
\begin{equation}
    \mathcal{N}_{\Psi_0} = \frac{1}{L} \frac{\langle \Psi| C_L^{\dagger} T_{L+1} C_L|\Psi\rangle}{\langle \Psi|\Psi\rangle},
\end{equation}
where we used $T_L^2 = L T_L$ and due to lack of the translational symmetry $\langle \Psi | T_L |\Psi\rangle = \langle \Psi |\Psi\rangle$.

{\bf Bound on the product states of zero magnetization:} 
Consider $|\Psi\rangle$ of zero magnetization (an eigenstate of the total $S_z$ operator with a zero eigenvalue) and a product state. We will now show that then $\mathcal{N} \geq 2$. Given the above bound it is enough to prove it for seed states without translational symmetry. Therefore, we want to show that  
\begin{equation} \label{inequality}
    \langle \Psi | C_L^{\dagger} T_L C_L | \Psi\rangle \geq 2 L \langle \Psi |\Psi\rangle.
\end{equation}
Because $|\Psi\rangle$ is a product state of zero magnetization it always contains at least one pair of adjacent spins $\downarrow \uparrow$.
Given the invariance of $\mathcal{N}$ under translations of the seed state we can always choose $|\Psi\rangle$ such that it starts with the spin up and ends with the spin down. From the zero magnetization condition we also know that there are exactly $L/2$ sites with spin ups. The term on the left hand side of the equality can be divided into two contributions
\begin{equation}
    \langle \Psi | C_L^{\dagger} T_{L+1} C_L | \Psi\rangle =  \langle \Psi | C_L^{\dagger} C_L | \Psi\rangle + \langle \Psi | C_L^{\dagger} (T_{L+1} - 1)C_L | \Psi\rangle.
\end{equation}
From the construction we know that the two contributions are also non-negative. We will show that already the first contribution itself saturates the bound, 
\begin{equation} \label{inequality_1}
    \langle \Psi | C_L^{\dagger} C_L | \Psi\rangle = 2 L \langle \Psi |\Psi\rangle.
\end{equation}
We need the prove it only for states with no translational invariance but it is easier to consider any product state of zero magnetization. 

\begin{figure}
    \center
    \includegraphics[scale=0.5]{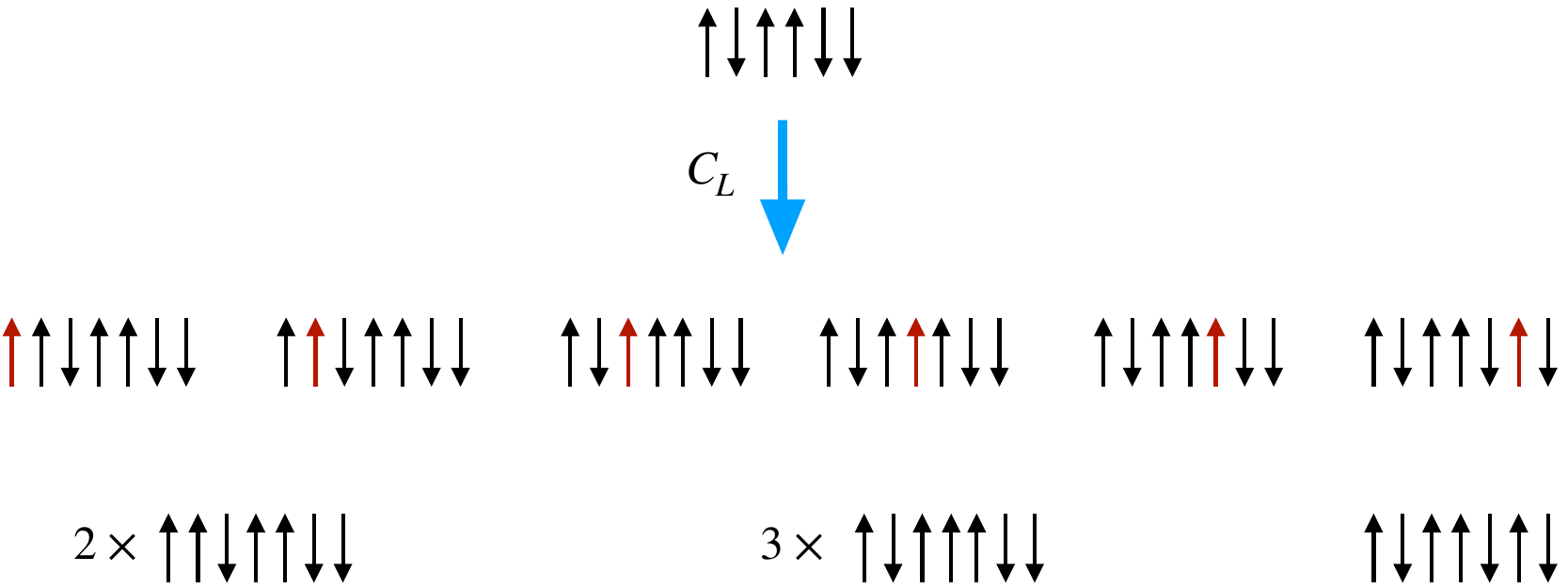}
    \label{fig:entanglement}
    \caption{Illustration of contributions to $\langle \Psi | C_L^{\dagger} C_L | \Psi\rangle$. For $|\Psi\rangle$ we choose a zero magnetization state without any translational symmetry. Action of $C_L$ inserts spin up (in red) at $6$ different positions leading to a superposition of $3$ different states. Value of $\langle \Psi | C_L^{\dagger} C_L | \Psi\rangle$ is then directly related to the multiplicity of those states. In the example given we find $\langle \Psi | C_L^{\dagger} C_L | \Psi\rangle = 2^2 + 3^2 + 1 = 14 \geq 2 L$ taking $\langle \Psi | \Psi\rangle = 1$. The multiplicities are directly related to the lengths of the domains of spin ups present in the seed state. Denoting by $m$ the length of a domain to which we add a spin, the multiplicity of the resulting state is $m+1$. This is valid also for $m=0$ when we insert the spin up between the spin downs as on the right hand side of the figure.}
\end{figure}

The idea of the proof is shown in fig.~\ref{fig:entanglement}. A generic product state contains domains of spins up and down of different lengths. Action of $C_L$ leads then to a superposition of new states which occur with multiplicities directly related to the structure of the domains of spin ups. For a state $|\Psi\rangle$ which contains a set of domains of spin up of sizes $\{m_i\}$ we then have
\begin{equation}
    \langle \Psi | C_L^{\dagger} C_L | \Psi\rangle = \sum_i (m_i+1)^2,
\end{equation}
because if by adding a spin up we enlarge a domain of length $m$ this can be achieved in $m+1$ ways and therefore contributes $(m+1)^2$ to the norm. For a state of zero magnetization we have a constraint $\sum_i m_i = L/2$. Minimizing the right hand side under this constraint leads to $\bar{m}_i = 1$ for $i = 1, \dots, L/2$ for which $\sum_i (\bar{m}_i+1)^2 = 2L$ which proves the equality~\eqref{inequality_1} and in turn inequality~\eqref{inequality}. 

This completes a proof that $\mathcal{N}\geq 2$ for product states of zero magnetization.

{\bf Neel state and its generalizations:}

Let us exemplify these concepts with direct computations for a simple case of a Neel state. The translationally invariant state is the combination
\begin{equation} \label{app_Neel}
    |\Psi_0\rangle = |\!\uparrow \downarrow \uparrow \downarrow \cdots \rangle + |\!\downarrow \uparrow \downarrow \uparrow \cdots \rangle,
\end{equation}
and, up to a normalization, it can be generated by the action of $T_L$ on one of the two seed states, $|\!\uparrow \downarrow \uparrow \downarrow \cdots \rangle$ or $|\!\downarrow \uparrow \downarrow \uparrow \cdots \rangle$. By explicit computations we find $N_{\Psi_0} = 2$. The Neel state has a partial translational invariance with the elementary cell given by either $|\!\uparrow \downarrow\rangle$ or $|\!\downarrow \uparrow\rangle$ for which we find $N_{\psi_0} = 2$. In the case of the Neel state the bound~\eqref{app_bound_cell} is actually saturated. The Neel state saturates also the bound $N_{\Psi_0} \geq 2$ for the product states of zero magnetization.

The Neel state can be generalized by taking alternatively domains of $m_1$ spin ups and $m_2$ spin downs. For a translationally invariant superposition of such states we find $\mathcal{N}_{m_1, m_2} = (m_1^2 + 2m_1 + m_2)/(m_1 + m_2)$. When $m_1 \neq m_2$ the state has finite magnetization and $N_{m_1, m_2}$ can be smaller than $2$ --- for example by taking $m_2$ much larger than $m_1$. However when $m = m_1 = m_2$ (with the obvious condition $m\geq 1$) we find $\mathcal{N}_{m, m} = (m + 3)/2 \geq 2$ with equality only for $m=1$ which is again the Neel state.

\section{Adding a single spin to other 1D spin model ground states}

\begin{figure}
    \center
    \includegraphics[scale=0.75]{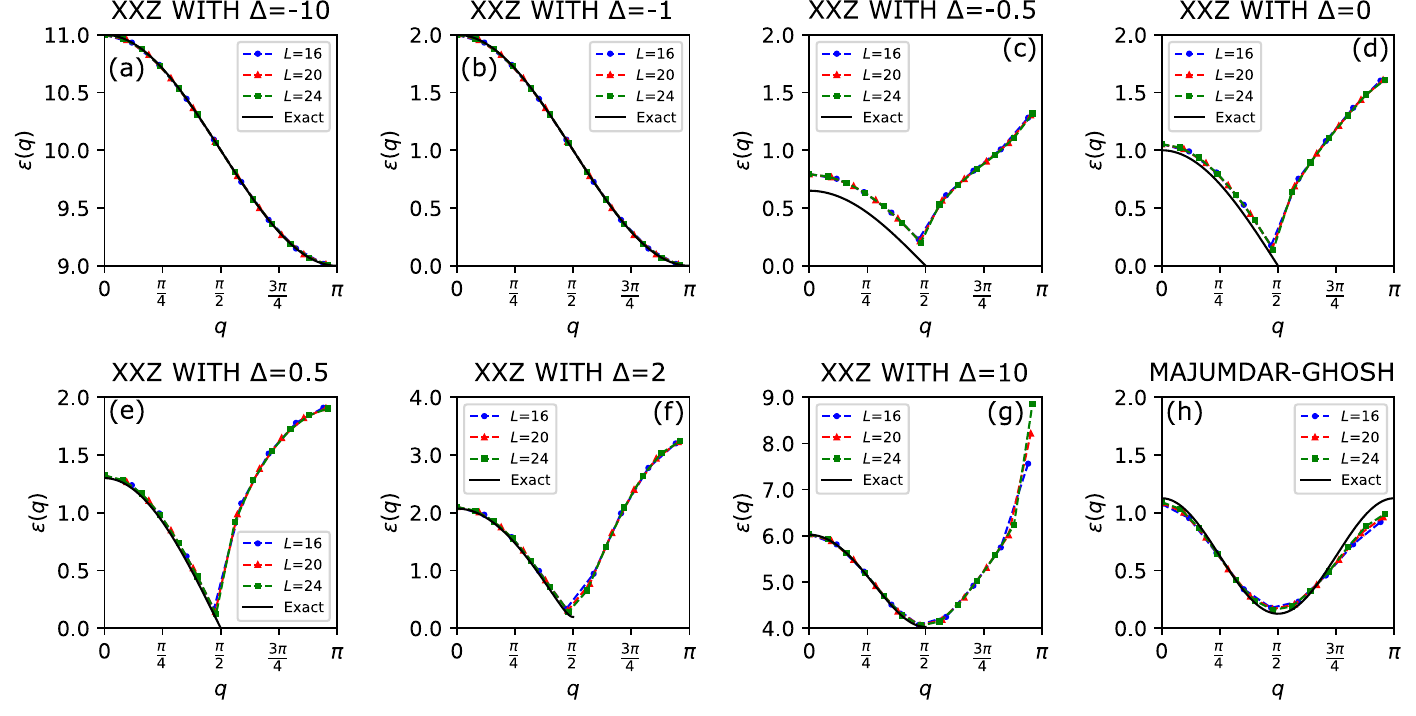}
    \caption{Dispersion relation $\varepsilon (q)$ of the excitation calculated using the same procedure as  discussed in the main text of the paper for: (a-g) the XXZ model \eqref{eq:xxz} with different values of the anisotropy $\Delta$, (h) the Majumdar-Ghosh model  \eqref{eq:mg} that has the valence-bond solid as the ground state. Results for chains of various length $L$ are shown by symbols (the connecting dashed lines are guides to the eye). The thick black solid lines, for comparison purposes, show the exact analytical expressions: (a-g) for the XXZ model originating in the Bethe Ansatz, and (h) for the Majumdar-Ghosh model, see text for more details.
    }
    \label{fig:othermodels}
\end{figure}

Here we show that the procedure, discussed in the main text works well also for other 1D spin models. [We note in passing a (successful) application of the procedure to the Shastry-Sutherland model discussed in \cite{Talstra1997}]. To this end, we consider two models: 

The first is the XXZ model defined by
\begin{align} \label{eq:xxz}
	H^{\rm XXZ}_L = J \sum_{j=1}^L 
 \left( \sigma^x_j \cdot \sigma^x_{j+1} +  \sigma^y_j \cdot \sigma^y_{j+1}  
 +\Delta  \sigma^z_j \cdot \sigma^z_{j+1}
 \right),
\end{align}
where we set $J=1$ and {\it widely vary} the anisotropy $\Delta$ (cf. \cite{Talstra1997} for the XY model result). Note that the XXZ model with $\Delta = -1$ is equivalent to the FM Heisenberg model (the unitary transformation that relates these models shifts the wavevector of the magnon dispersion minimum).

The second is the Majumdar-Ghosh model with
its Hamiltonian 
\begin{align} \label{eq:mg}
	H^{\rm MG}_L = J \sum_{j=1}^L 
 \left( \vec{\sigma}_j \cdot \vec{\sigma}_{j+1} + \frac12 \vec{\sigma}_j \cdot \vec{\sigma}_{j+2} \right),
\end{align}
where again we set $J=1$.  

The corresponding dispersions $\varepsilon(q)$ obtained through the procedure proposed in the main text,  are shown in Fig.~\ref{fig:othermodels}.
For the XXZ model with easy axis anisotropy we observe perfect agreement with the results obtained using the Bethe Ansatz~\cite{Caux_2008,2017LNP...940.....F,Bohrdt2018,PhysRevB.106.134405}.
The agreement is quantitatively worse for the XXZ model with an easy plane anisotropy: In this case
more than one spinon is excited when a single spin is inserted in the ground state, cf. the XY limit discussed in~\cite{Talstra1997}.
For the Majumdar-Ghosh model only small deviations are observed from the exact dispersion, see Ref.~\cite{Shastry1981, Lavarelo2014}.

As a side note, in the Majumdar-Ghosh model  the dispersion scales
as $\propto q ^2$ in the low-energy limit~\cite{Shastry1981, Bohrdt2018},  despite having a ground state with valence bond character and thus the norm $N(q)$ nearly vanishing in an extended region of $|q| > \pi /2 $, just as in Fig.~2(c2) of the main text. The conundrum is resolved by noting that in the Majumdar-Ghosh Hamiltonian~\cite{Shastry1981} there exist next-nearest-neighbor spin flip terms which lead to a $\propto \cos (2q)$ contribution to the dispersion relation. 
%

{\section{Pump-probe experimental techniques that can excite a single spinon}}

We postulate that the following `pump-probe' style experiment should be able to measure a single spinon following the procedure outlined in the paper. While as of now it can be seen merely as a `Gedanken' experiment, it is conceivable that it can be realised in the future. The main concept behind this experiment is related to removing a single spin from a magnetic lattice and enhancing the spin exchange across the empty site in such away that it is {\it indistinguishable} from the spin exchange between the intact bonds -- in short we just `kick a spin and then glue the spins together'. Note that this procedure is equivalent to removing a single spin site and `contracting' the lattice -- yet it does not require changing the geometry of the lattice and hence not only it is valid in any dimension but also should not so easily excite phonons.

\begin{figure}
\centering
\includegraphics[width=0.95\linewidth]{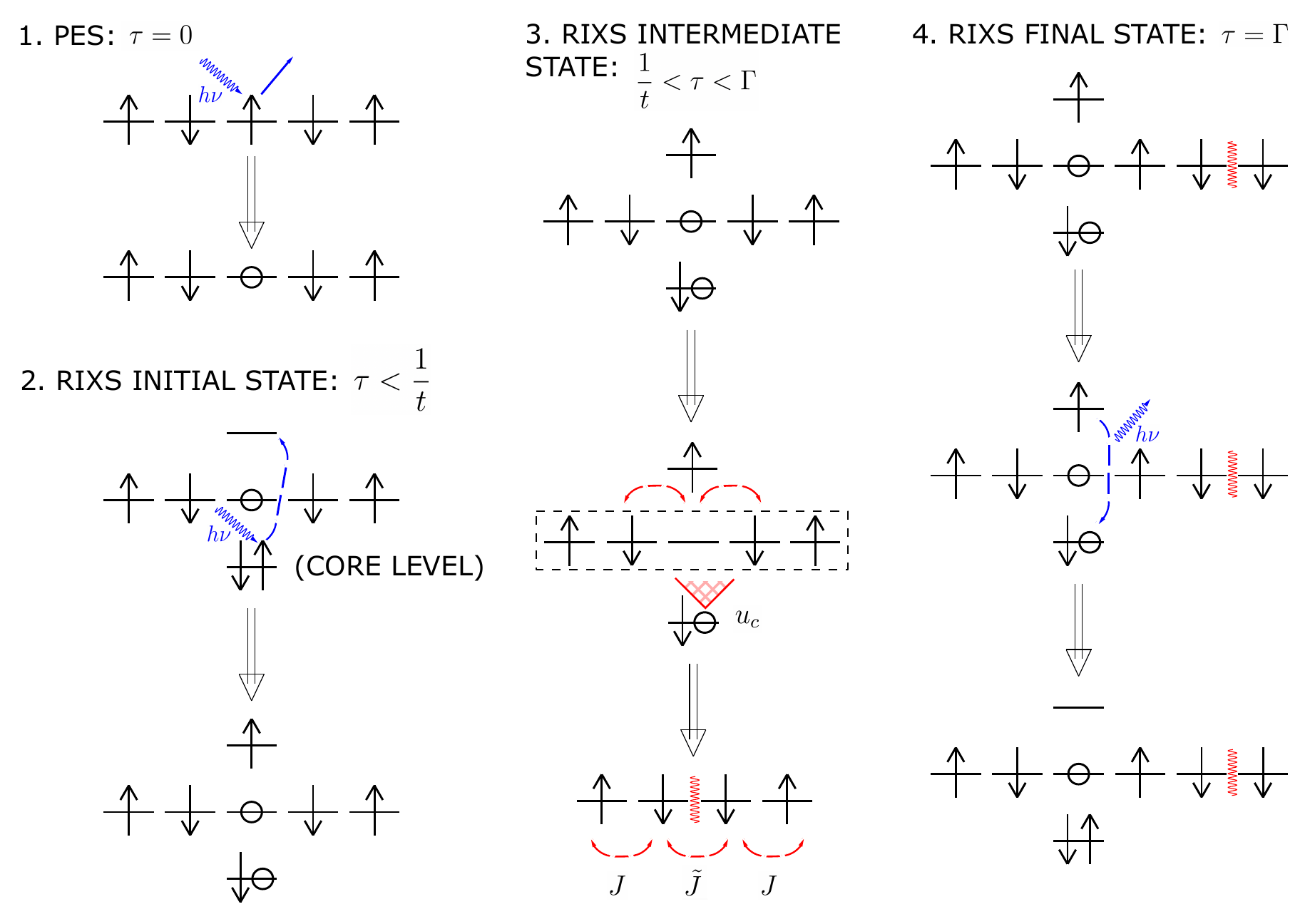}
\caption{
Cartoon of the pump and probe experiment that can excite a single spinon by `kicking a spin and gluing the spins together', i.e.
following a procedure that qualitatively mimics the one outlined in the main text of the paper:
{\it Left panel -- steps 1 and 2}: photoemission spectroscopy (PES) creates a single hole
that is followed by a single core-hole created by the resonant-inelastic x-ray scattering (RIXS) at the same site and at time $\tau \ll 1/t$; {\it Middle panel -- step 3}: 
during time $0<\tau < \Gamma$ the RIXS core-hole potential lowers the value to such an extent
that the value of the spin exchange $\tilde{J}$ across
the empty site is equal to the exchange between nearest neighbor spins (see text for further details);
{\it Right panel -- step 4}: At time $\tau \sim \Gamma$
the RIXS photon is emitted, its energy and momentum is measured, and the core-hole filled.
Note that in this simplified cartoon picture the pumps and the probe are aligned at the same lattice site $j$ -- while in reality the two are aligned at the same momentum $k$ (see text for further details). 
}
\label{fig:cartoon}
\end{figure}

In detail, such an experiment can be designed in the following manner:

\begin{enumerate}
    \item At time $\tau=0$ and momentum $k$ in the 1st Brillouin zone a photoemission experiment
    (PES) creates a hole in the valence shell (i.e. one spin is kicked out with a well-defined momentum from the valence shell). This pump and its action is illustrated in left panel of Fig.~\ref{fig:cartoon}.
 \item At time $\tau \ll 1/t$ and momentum $k$ in the 1st Brillouin zone an indirect resonant inelastic x-ray scattering (RIXS) experiment (e.g. $K$ edge) creates a core hole. This way simultaneously with creating a hole in the spin lattice, a core-hole potential is created at the same site (in momentum space), see left panel of Fig.~\ref{fig:cartoon}.
 \item Next, at time $\tau \in [1/t, \Gamma]$, where $\Gamma$ is the core-hole lifetime, the system can create a single spinon excitation, provided the core-hole potential $U_c$ has a particular value w.r.t $U$: 
 
 For instance, if we assume that  
 $U\sim 8t$ (a rather standard value for a Hubbard model that yields the typical spin exchange $J=0.5t$) {\it and}
 that $U_c \sim U/3$ then we obtain that
the spin  exchange $\tilde{J}$ between the spins across the empty site~\footnote{Here we apply the standard superexchange formula, [{\it cf}.~\cite{Zaanen1988} or Eq. (7) in \cite{Wohlfeld2010}] to the bond with three transition metal ions.} reads
 \begin{equation}
     \tilde{J}= 4 \frac{t^4}{\Delta^2} \left( \frac{2}{2 \Delta + U} + \frac{1}{U} \right)\Big|_{\Delta = -U_c} 
     = 4 \frac{t^4}{U_c^2} \left( \frac{2}{ U-2U_c} + \frac{1}{U} \right) \Big|_{U_c = U/3\ \&\ U= 8t} \sim 0.5t,
 \end{equation}
 and hence it is equal to the `canonical' spin exchange $J = 0.5t$ between all `other' spins, see middle panel of Fig.~\ref{fig:cartoon}. Thus, we create conditions that resemble the procedure postulated in the main text of the paper -- but `\`{a} rebours', i.e. effectively we end up with a removed spin site. This is because the spin exchange {\it across} the empty site $\tilde{J}$ is the same as the spin exchange $J$ elsewhere.
\item Finally, at time $\tau=\Gamma$ the RIXS x-ray is emitted and its momentum and energy measured, see right panel of Fig.~\ref{fig:cartoon}. As the energy and momentum lost by RIXS is taken by the single spinon excited by RIXS, we can this way measure the single spinon dispersion.
\end{enumerate}

Note that the above procedure does not depend on dimensionality.
However, it requires finding a spin liquid based on a single-band Hubbard and that possesses a fine-tuned
core-hole potential $U_c$ at the chosen indirect RIXS edge (although overall its value is reasonable~\cite{Ament2011, Jia2016}).
Finally, both pumps have to be perfectly aligned in time with the RIXS probe (incl. the core-hole lifetime).

We may wonder whether this way we can indeed excite just single spinons in higher dimensions -- since removing a single site from e.g. a 2D square lattice also locally changes the topology of the lattice (which can be probed e.g. by traversing a loop around the created vacancy). 
While we leave this question to future studies, 
the point is that the presented scheme clearly excites
single spinons in 1D and  hence one should investigate in the future whether its natural 
generalisation to higher dimensions
also predominantly excites single spinons -- 
or there are also other excitations involved.

\begin{figure}
\centering
\includegraphics[width=0.45\linewidth]{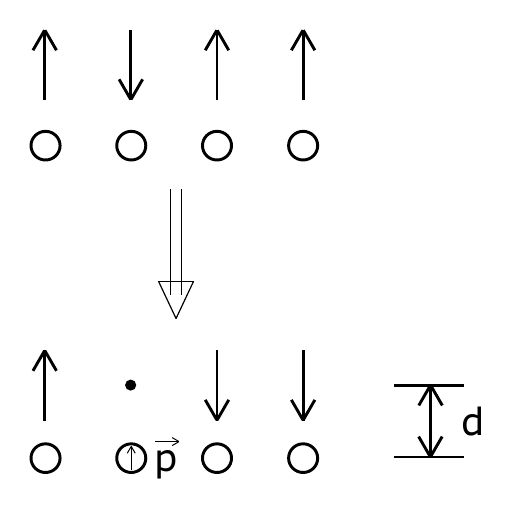}
\caption{
Cartoon
showing how one can `kick a spin and glue the spins together' using the photoemission and the polarisable ions:
Spin lattice and the  mirrored lattice that is populated by polarizable ions (top panel); When one of the spins is removed by the photomeission process, the resulting uncompensated charge at that site creates a local electric field that polarizes the nearby polarizable ions and lowers the total energy~\cite{Sawatzky2009} (bottom panel). Note that the latter can be set to the value needed to ensure that spin exchange $\tilde{J}$ across the empty site is equal to the exchange between nearest-neighbour spins (see text for further details). 
}
\label{fig:cartoonpolar}
\end{figure}

{\it Last but not least}, we note that instead of using RIXS, an alternative way to generate the coupling $\tilde{J} \sim J$ across the emptied site is to use the polarizability of ions, as suggested in 
Fig~\ref{fig:cartoonpolar}. For simplicity of drawing, the panels illustrate the idea for a 1D chain, but it can be generalized straightforwardly to higher dimensions.

In this procedure, we assume that the original spin lattice is mirrored by another similar lattice, populated by polarizable ions with full atomic shells and a polarizability $\alpha$ (top panel). When one of the spins is removed by the PES process (bottom panel), the resulting uncompensated charge at that site creates a local electric field that polarizes the nearby polarizable ions. This lowers the total energy by an amount of order $\sum_{\delta} \alpha {\vec E(\delta)}^2/2$, where $\vec E(\delta)$ is the electric field of the uncompensated charge at the polarizable ion located $\delta$ sites away~\cite{Sawatzky2009}. As such, this energy can be controlled by the appropriate choice of the polarizable ions and of their positioning with respect to the spin lattice. In particular, it can be set to the value needed to ensure that spin exchange $\tilde{J}$ across the empty site is equal to the exchange between nearest-neighbour spins.

\begin{thebibliography}{45}%
\makeatletter
\providecommand \@ifxundefined [1]{%
 \@ifx{#1\undefined}
}%
\providecommand \@ifnum [1]{%
 \ifnum #1\expandafter \@firstoftwo
 \else \expandafter \@secondoftwo
 \fi
}%
\providecommand \@ifx [1]{%
 \ifx #1\expandafter \@firstoftwo
 \else \expandafter \@secondoftwo
 \fi
}%
\providecommand \natexlab [1]{#1}%
\providecommand \enquote  [1]{``#1''}%
\providecommand \bibnamefont  [1]{#1}%
\providecommand \bibfnamefont [1]{#1}%
\providecommand \citenamefont [1]{#1}%
\providecommand \href@noop [0]{\@secondoftwo}%
\providecommand \href [0]{\begingroup \@sanitize@url \@href}%
\providecommand \@href[1]{\@@startlink{#1}\@@href}%
\providecommand \@@href[1]{\endgroup#1\@@endlink}%
\providecommand \@sanitize@url [0]{\catcode `\\12\catcode `\$12\catcode
  `\&12\catcode `\#12\catcode `\^12\catcode `\_12\catcode `\%12\relax}%
\providecommand \@@startlink[1]{}%
\providecommand \@@endlink[0]{}%
\providecommand \url  [0]{\begingroup\@sanitize@url \@url }%
\providecommand \@url [1]{\endgroup\@href {#1}{\urlprefix }}%
\providecommand \urlprefix  [0]{URL }%
\providecommand \Eprint [0]{\href }%
\providecommand \doibase [0]{http://dx.doi.org/}%
\providecommand \selectlanguage [0]{\@gobble}%
\providecommand \bibinfo  [0]{\@secondoftwo}%
\providecommand \bibfield  [0]{\@secondoftwo}%
\providecommand \translation [1]{[#1]}%
\providecommand \BibitemOpen [0]{}%
\providecommand \bibitemStop [0]{}%
\providecommand \bibitemNoStop [0]{.\EOS\space}%
\providecommand \EOS [0]{\spacefactor3000\relax}%
\providecommand \BibitemShut  [1]{\csname bibitem#1\endcsname}%
\let\auto@bib@innerbib\@empty
\bibitem [{\citenamefont {Khomskii}(2010)}]{Khomskii2010}%
  \BibitemOpen
  \bibfield  {author} {\bibinfo {author} {\bibfnamefont {D.~I.}\ \bibnamefont
  {Khomskii}},\ }\href@noop {} {\emph {\bibinfo {title} {Basic Aspects of the
  Quantum Theory of Solids: Order and Elementary Excitations}}}\ (\bibinfo
  {publisher} {Cambridge University Press},\ \bibinfo {address} {Cambridge},\
  \bibinfo {year} {2010})\BibitemShut {NoStop}%
\bibitem [{\citenamefont {Venema}\ \emph {et~al.}(2016)\citenamefont {Venema},
  \citenamefont {Verberck}, \citenamefont {Georgescu}, \citenamefont {Prando},
  \citenamefont {Couderc}, \citenamefont {Milana}, \citenamefont {Maragkou},
  \citenamefont {Persechini}, \citenamefont {Pacchioni},\ and\ \citenamefont
  {Fleet}}]{Venema2016}%
  \BibitemOpen
  \bibfield  {author} {\bibinfo {author} {\bibfnamefont {L.}~\bibnamefont
  {Venema}}, \bibinfo {author} {\bibfnamefont {B.}~\bibnamefont {Verberck}},
  \bibinfo {author} {\bibfnamefont {I.}~\bibnamefont {Georgescu}}, \bibinfo
  {author} {\bibfnamefont {G.}~\bibnamefont {Prando}}, \bibinfo {author}
  {\bibfnamefont {E.}~\bibnamefont {Couderc}}, \bibinfo {author} {\bibfnamefont
  {S.}~\bibnamefont {Milana}}, \bibinfo {author} {\bibfnamefont
  {M.}~\bibnamefont {Maragkou}}, \bibinfo {author} {\bibfnamefont
  {L.}~\bibnamefont {Persechini}}, \bibinfo {author} {\bibfnamefont
  {G.}~\bibnamefont {Pacchioni}}, \ and\ \bibinfo {author} {\bibfnamefont
  {L.}~\bibnamefont {Fleet}},\ }\href {\doibase 10.1038/nphys3977} {\bibfield
  {journal} {\bibinfo  {journal} {Nat. Phys.}\ }\textbf {\bibinfo {volume}
  {12}},\ \bibinfo {pages} {1085} (\bibinfo {year} {2016})}\BibitemShut
  {NoStop}%
\bibitem [{\citenamefont {Auerbach}(1994)}]{Auerbach1994}%
  \BibitemOpen
  \bibfield  {author} {\bibinfo {author} {\bibfnamefont {A.}~\bibnamefont
  {Auerbach}},\ }\href@noop {} {\emph {\bibinfo {title} {Interacting Electrons
  and Quantum Magnetism}}}\ (\bibinfo  {publisher} {Springer-Verlag, New
  York},\ \bibinfo {year} {1994})\BibitemShut {NoStop}%
\bibitem [{\citenamefont {Bethe}(1931)}]{1931_Bethe_ZP_71}%
  \BibitemOpen
  \bibfield  {author} {\bibinfo {author} {\bibfnamefont {H.}~\bibnamefont
  {Bethe}},\ }\href@noop {} {\bibfield  {journal} {\bibinfo  {journal} {Zeit.
  f\"ur Physik}\ }\textbf {\bibinfo {volume} {71}},\ \bibinfo {pages} {205}
  (\bibinfo {year} {1931})}\BibitemShut {NoStop}%
\bibitem [{\citenamefont {Faddeev}\ and\ \citenamefont
  {Takhtajan}(1981)}]{1981_Faddeev_PLA_85}%
  \BibitemOpen
  \bibfield  {author} {\bibinfo {author} {\bibfnamefont {L.~D.}\ \bibnamefont
  {Faddeev}}\ and\ \bibinfo {author} {\bibfnamefont {L.~A.}\ \bibnamefont
  {Takhtajan}},\ }\href@noop {} {\bibfield  {journal} {\bibinfo  {journal}
  {Phys. Lett. A}\ }\textbf {\bibinfo {volume} {85}},\ \bibinfo {pages} {375}
  (\bibinfo {year} {1981})}\BibitemShut {NoStop}%
\bibitem [{\citenamefont {des Cloizeaux}\ and\ \citenamefont
  {Pearson}(1962)}]{1962_desCloizeaux_PR_128}%
  \BibitemOpen
  \bibfield  {author} {\bibinfo {author} {\bibfnamefont {J.}~\bibnamefont {des
  Cloizeaux}}\ and\ \bibinfo {author} {\bibfnamefont {J.~J.}\ \bibnamefont
  {Pearson}},\ }\href {\doibase 10.1103/PhysRev.128.2131} {\bibfield  {journal}
  {\bibinfo  {journal} {Phys. Rev.}\ }\textbf {\bibinfo {volume} {128}},\
  \bibinfo {pages} {2131} (\bibinfo {year} {1962})}\BibitemShut {NoStop}%
\bibitem [{\citenamefont {Tennant}\ \emph {et~al.}(1993)\citenamefont
  {Tennant}, \citenamefont {Perring}, \citenamefont {Cowley},\ and\
  \citenamefont {Nagler}}]{Tennant1993}%
  \BibitemOpen
  \bibfield  {author} {\bibinfo {author} {\bibfnamefont {D.~A.}\ \bibnamefont
  {Tennant}}, \bibinfo {author} {\bibfnamefont {T.~G.}\ \bibnamefont
  {Perring}}, \bibinfo {author} {\bibfnamefont {R.~A.}\ \bibnamefont {Cowley}},
  \ and\ \bibinfo {author} {\bibfnamefont {S.~E.}\ \bibnamefont {Nagler}},\
  }\href {\doibase 10.1103/PhysRevLett.70.4003} {\bibfield  {journal} {\bibinfo
   {journal} {Phys. Rev. Lett.}\ }\textbf {\bibinfo {volume} {70}},\ \bibinfo
  {pages} {4003} (\bibinfo {year} {1993})}\BibitemShut {NoStop}%
\bibitem [{\citenamefont {Lake}\ \emph {et~al.}(2005)\citenamefont {Lake},
  \citenamefont {Tennant}, \citenamefont {Frost},\ and\ \citenamefont
  {Nagler}}]{Lake2005}%
  \BibitemOpen
  \bibfield  {author} {\bibinfo {author} {\bibfnamefont {B.}~\bibnamefont
  {Lake}}, \bibinfo {author} {\bibfnamefont {D.~A.}\ \bibnamefont {Tennant}},
  \bibinfo {author} {\bibfnamefont {C.~D.}\ \bibnamefont {Frost}}, \ and\
  \bibinfo {author} {\bibfnamefont {S.~E.}\ \bibnamefont {Nagler}},\ }\href
  {\doibase 10.1038/nmat1327} {\bibfield  {journal} {\bibinfo  {journal}
  {Nature Materials}\ }\textbf {\bibinfo {volume} {4}},\ \bibinfo {pages} {329}
  (\bibinfo {year} {2005})}\BibitemShut {NoStop}%
\bibitem [{\citenamefont {Caux}\ and\ \citenamefont
  {Maillet}(2005)}]{Caux2005}%
  \BibitemOpen
  \bibfield  {author} {\bibinfo {author} {\bibfnamefont {J.-S.}\ \bibnamefont
  {Caux}}\ and\ \bibinfo {author} {\bibfnamefont {J.~M.}\ \bibnamefont
  {Maillet}},\ }\href {\doibase 10.1103/PhysRevLett.95.077201} {\bibfield
  {journal} {\bibinfo  {journal} {Phys. Rev. Lett.}\ }\textbf {\bibinfo
  {volume} {95}},\ \bibinfo {pages} {077201} (\bibinfo {year}
  {2005})}\BibitemShut {NoStop}%
\bibitem [{\citenamefont {Klauser}\ \emph {et~al.}(2011)\citenamefont
  {Klauser}, \citenamefont {Mossel}, \citenamefont {Caux},\ and\ \citenamefont
  {van~den Brink}}]{Klauser2011}%
  \BibitemOpen
  \bibfield  {author} {\bibinfo {author} {\bibfnamefont {A.}~\bibnamefont
  {Klauser}}, \bibinfo {author} {\bibfnamefont {J.}~\bibnamefont {Mossel}},
  \bibinfo {author} {\bibfnamefont {J.-S.}\ \bibnamefont {Caux}}, \ and\
  \bibinfo {author} {\bibfnamefont {J.}~\bibnamefont {van~den Brink}},\ }\href
  {\doibase 10.1103/PhysRevLett.106.157205} {\bibfield  {journal} {\bibinfo
  {journal} {Phys. Rev. Lett.}\ }\textbf {\bibinfo {volume} {106}},\ \bibinfo
  {pages} {157205} (\bibinfo {year} {2011})}\BibitemShut {NoStop}%
\bibitem [{\citenamefont {Schlappa}\ \emph {et~al.}(2012)\citenamefont
  {Schlappa}, \citenamefont {Wohlfeld}, \citenamefont {Zhou}, \citenamefont
  {Mourigal}, \citenamefont {Haverkort}, \citenamefont {Strocov}, \citenamefont
  {Hozoi}, \citenamefont {Monney}, \citenamefont {Nishimoto}, \citenamefont
  {Singh}, \citenamefont {Revcolevschi}, \citenamefont {Caux}, \citenamefont
  {Patthey}, \citenamefont {R{\o}nnow}, \citenamefont {van~den Brink},\ and\
  \citenamefont {Schmitt}}]{Schlappa2012}%
  \BibitemOpen
  \bibfield  {author} {\bibinfo {author} {\bibfnamefont {J.}~\bibnamefont
  {Schlappa}}, \bibinfo {author} {\bibfnamefont {K.}~\bibnamefont {Wohlfeld}},
  \bibinfo {author} {\bibfnamefont {K.~J.}\ \bibnamefont {Zhou}}, \bibinfo
  {author} {\bibfnamefont {M.}~\bibnamefont {Mourigal}}, \bibinfo {author}
  {\bibfnamefont {M.~W.}\ \bibnamefont {Haverkort}}, \bibinfo {author}
  {\bibfnamefont {V.~N.}\ \bibnamefont {Strocov}}, \bibinfo {author}
  {\bibfnamefont {L.}~\bibnamefont {Hozoi}}, \bibinfo {author} {\bibfnamefont
  {C.}~\bibnamefont {Monney}}, \bibinfo {author} {\bibfnamefont
  {S.}~\bibnamefont {Nishimoto}}, \bibinfo {author} {\bibfnamefont
  {S.}~\bibnamefont {Singh}}, \bibinfo {author} {\bibfnamefont
  {A.}~\bibnamefont {Revcolevschi}}, \bibinfo {author} {\bibfnamefont {J.-S.}\
  \bibnamefont {Caux}}, \bibinfo {author} {\bibfnamefont {L.}~\bibnamefont
  {Patthey}}, \bibinfo {author} {\bibfnamefont {H.~M.}\ \bibnamefont
  {R{\o}nnow}}, \bibinfo {author} {\bibfnamefont {J.}~\bibnamefont {van~den
  Brink}}, \ and\ \bibinfo {author} {\bibfnamefont {T.}~\bibnamefont
  {Schmitt}},\ }\href {\doibase 10.1038/nature10974} {\bibfield  {journal}
  {\bibinfo  {journal} {Nature}\ }\textbf {\bibinfo {volume} {485}},\ \bibinfo
  {pages} {82} (\bibinfo {year} {2012})}\BibitemShut {NoStop}%
\bibitem [{\citenamefont {Mourigal}\ \emph {et~al.}(2013)\citenamefont
  {Mourigal}, \citenamefont {Enderle}, \citenamefont {Kl\"{o}pperpieper},
  \citenamefont {Caux}, \citenamefont {Stunault},\ and\ \citenamefont
  {R{\o}nnow}}]{Mourigal2013}%
  \BibitemOpen
  \bibfield  {author} {\bibinfo {author} {\bibfnamefont {M.}~\bibnamefont
  {Mourigal}}, \bibinfo {author} {\bibfnamefont {M.}~\bibnamefont {Enderle}},
  \bibinfo {author} {\bibfnamefont {A.}~\bibnamefont {Kl\"{o}pperpieper}},
  \bibinfo {author} {\bibfnamefont {J.-S.}\ \bibnamefont {Caux}}, \bibinfo
  {author} {\bibfnamefont {A.}~\bibnamefont {Stunault}}, \ and\ \bibinfo
  {author} {\bibfnamefont {H.~M.}\ \bibnamefont {R{\o}nnow}},\ }\href {\doibase
  10.1038/nphys2652} {\bibfield  {journal} {\bibinfo  {journal} {Nature
  Physics}\ }\textbf {\bibinfo {volume} {9}},\ \bibinfo {pages} {435} (\bibinfo
  {year} {2013})}\BibitemShut {NoStop}%
\bibitem [{\citenamefont {Ferrari}\ \emph {et~al.}(2018)\citenamefont
  {Ferrari}, \citenamefont {Parola}, \citenamefont {Sorella},\ and\
  \citenamefont {Becca}}]{Ferrari2018}%
  \BibitemOpen
  \bibfield  {author} {\bibinfo {author} {\bibfnamefont {F.}~\bibnamefont
  {Ferrari}}, \bibinfo {author} {\bibfnamefont {A.}~\bibnamefont {Parola}},
  \bibinfo {author} {\bibfnamefont {S.}~\bibnamefont {Sorella}}, \ and\
  \bibinfo {author} {\bibfnamefont {F.}~\bibnamefont {Becca}},\ }\href
  {\doibase 10.1103/PhysRevB.97.235103} {\bibfield  {journal} {\bibinfo
  {journal} {Phys. Rev. B}\ }\textbf {\bibinfo {volume} {97}},\ \bibinfo
  {pages} {235103} (\bibinfo {year} {2018})}\BibitemShut {NoStop}%
\bibitem [{\citenamefont {Sorella}\ and\ \citenamefont
  {Parola}(1992)}]{Sorella1992}%
  \BibitemOpen
  \bibfield  {author} {\bibinfo {author} {\bibfnamefont {S.}~\bibnamefont
  {Sorella}}\ and\ \bibinfo {author} {\bibfnamefont {A.}~\bibnamefont
  {Parola}},\ }\href {\doibase 10.1088/0953-8984/4/13/020} {\bibfield
  {journal} {\bibinfo  {journal} {Journal of Physics: Condensed Matter}\
  }\textbf {\bibinfo {volume} {4}},\ \bibinfo {pages} {3589} (\bibinfo {year}
  {1992})}\BibitemShut {NoStop}%
\bibitem [{\citenamefont {Talstra}\ and\ \citenamefont
  {Strong}(1997)}]{Talstra1997}%
  \BibitemOpen
  \bibfield  {author} {\bibinfo {author} {\bibfnamefont {J.~C.}\ \bibnamefont
  {Talstra}}\ and\ \bibinfo {author} {\bibfnamefont {S.~P.}\ \bibnamefont
  {Strong}},\ }\href {\doibase 10.1103/PhysRevB.56.6094} {\bibfield  {journal}
  {\bibinfo  {journal} {Phys. Rev. B}\ }\textbf {\bibinfo {volume} {56}},\
  \bibinfo {pages} {6094} (\bibinfo {year} {1997})}\BibitemShut {NoStop}%
\bibitem [{\citenamefont {Penc}\ \emph {et~al.}(1997)\citenamefont {Penc},
  \citenamefont {Hallberg}, \citenamefont {Mila},\ and\ \citenamefont
  {Shiba}}]{Penc1997}%
  \BibitemOpen
  \bibfield  {author} {\bibinfo {author} {\bibfnamefont {K.}~\bibnamefont
  {Penc}}, \bibinfo {author} {\bibfnamefont {K.}~\bibnamefont {Hallberg}},
  \bibinfo {author} {\bibfnamefont {F.}~\bibnamefont {Mila}}, \ and\ \bibinfo
  {author} {\bibfnamefont {H.}~\bibnamefont {Shiba}},\ }\href {\doibase
  10.1103/PhysRevB.55.15475} {\bibfield  {journal} {\bibinfo  {journal} {Phys.
  Rev. B}\ }\textbf {\bibinfo {volume} {55}},\ \bibinfo {pages} {15475}
  (\bibinfo {year} {1997})}\BibitemShut {NoStop}%
\bibitem [{\citenamefont {Penc}\ and\ \citenamefont
  {Serhan}(1997)}]{Penc1997b}%
  \BibitemOpen
  \bibfield  {author} {\bibinfo {author} {\bibfnamefont {K.}~\bibnamefont
  {Penc}}\ and\ \bibinfo {author} {\bibfnamefont {M.}~\bibnamefont {Serhan}},\
  }\href {\doibase 10.1103/PhysRevB.56.6555} {\bibfield  {journal} {\bibinfo
  {journal} {Phys. Rev. B}\ }\textbf {\bibinfo {volume} {56}},\ \bibinfo
  {pages} {6555} (\bibinfo {year} {1997})}\BibitemShut {NoStop}%
\bibitem [{\citenamefont {Matveev}\ \emph
  {et~al.}(2007{\natexlab{a}})\citenamefont {Matveev}, \citenamefont
  {Furusaki},\ and\ \citenamefont {Glazman}}]{Matveev2007a}%
  \BibitemOpen
  \bibfield  {author} {\bibinfo {author} {\bibfnamefont {K.~A.}\ \bibnamefont
  {Matveev}}, \bibinfo {author} {\bibfnamefont {A.}~\bibnamefont {Furusaki}}, \
  and\ \bibinfo {author} {\bibfnamefont {L.~I.}\ \bibnamefont {Glazman}},\
  }\href {\doibase 10.1103/PhysRevLett.98.096403} {\bibfield  {journal}
  {\bibinfo  {journal} {Phys. Rev. Lett.}\ }\textbf {\bibinfo {volume} {98}},\
  \bibinfo {pages} {096403} (\bibinfo {year} {2007}{\natexlab{a}})}\BibitemShut
  {NoStop}%
\bibitem [{\citenamefont {Matveev}\ \emph
  {et~al.}(2007{\natexlab{b}})\citenamefont {Matveev}, \citenamefont
  {Furusaki},\ and\ \citenamefont {Glazman}}]{Matveev2007b}%
  \BibitemOpen
  \bibfield  {author} {\bibinfo {author} {\bibfnamefont {K.~A.}\ \bibnamefont
  {Matveev}}, \bibinfo {author} {\bibfnamefont {A.}~\bibnamefont {Furusaki}}, \
  and\ \bibinfo {author} {\bibfnamefont {L.~I.}\ \bibnamefont {Glazman}},\
  }\href {\doibase 10.1103/PhysRevB.76.155440} {\bibfield  {journal} {\bibinfo
  {journal} {Phys. Rev. B}\ }\textbf {\bibinfo {volume} {76}},\ \bibinfo
  {pages} {155440} (\bibinfo {year} {2007}{\natexlab{b}})}\BibitemShut
  {NoStop}%
\bibitem [{\citenamefont {Vidal}\ \emph
  {et~al.}(2003{\natexlab{a}})\citenamefont {Vidal}, \citenamefont {Latorre},
  \citenamefont {Rico},\ and\ \citenamefont {Kitaev}}]{PhysRevLett.90.227902}%
  \BibitemOpen
  \bibfield  {author} {\bibinfo {author} {\bibfnamefont {G.}~\bibnamefont
  {Vidal}}, \bibinfo {author} {\bibfnamefont {J.~I.}\ \bibnamefont {Latorre}},
  \bibinfo {author} {\bibfnamefont {E.}~\bibnamefont {Rico}}, \ and\ \bibinfo
  {author} {\bibfnamefont {A.}~\bibnamefont {Kitaev}},\ }\href {\doibase
  10.1103/PhysRevLett.90.227902} {\bibfield  {journal} {\bibinfo  {journal}
  {Phys. Rev. Lett.}\ }\textbf {\bibinfo {volume} {90}},\ \bibinfo {pages}
  {227902} (\bibinfo {year} {2003}{\natexlab{a}})}\BibitemShut {NoStop}%
\bibitem [{\citenamefont {Sato}\ \emph {et~al.}(2006)\citenamefont {Sato},
  \citenamefont {Shiroishi},\ and\ \citenamefont {Takahashi}}]{Sato_2006}%
  \BibitemOpen
  \bibfield  {author} {\bibinfo {author} {\bibfnamefont {J.}~\bibnamefont
  {Sato}}, \bibinfo {author} {\bibfnamefont {M.}~\bibnamefont {Shiroishi}}, \
  and\ \bibinfo {author} {\bibfnamefont {M.}~\bibnamefont {Takahashi}},\ }\href
  {\doibase 10.1088/1742-5468/2006/12/P12017} {\bibfield  {journal} {\bibinfo
  {journal} {Journal of Statistical Mechanics: Theory and Experiment}\ }\textbf
  {\bibinfo {volume} {2006}},\ \bibinfo {pages} {P12017} (\bibinfo {year}
  {2006})}\BibitemShut {NoStop}%
\bibitem [{\citenamefont {Miwa}\ and\ \citenamefont
  {Smirnov}(2019)}]{XXX_density_matrix}%
  \BibitemOpen
  \bibfield  {author} {\bibinfo {author} {\bibfnamefont {T.}~\bibnamefont
  {Miwa}}\ and\ \bibinfo {author} {\bibfnamefont {F.}~\bibnamefont {Smirnov}},\
  }\href {\doibase 10.1007/s11005-018-01143-x} {\bibfield  {journal} {\bibinfo
  {journal} {Letters in Mathematical Physics}\ }\textbf {\bibinfo {volume}
  {109}},\ \bibinfo {pages} {675} (\bibinfo {year} {2019})}\BibitemShut
  {NoStop}%
\bibitem [{\citenamefont {Murciano}\ \emph {et~al.}(2020)\citenamefont
  {Murciano}, \citenamefont {Giulio},\ and\ \citenamefont
  {Calabrese}}]{10.21468/SciPostPhys.8.3.046}%
  \BibitemOpen
  \bibfield  {author} {\bibinfo {author} {\bibfnamefont {S.}~\bibnamefont
  {Murciano}}, \bibinfo {author} {\bibfnamefont {G.~D.}\ \bibnamefont
  {Giulio}}, \ and\ \bibinfo {author} {\bibfnamefont {P.}~\bibnamefont
  {Calabrese}},\ }\href {\doibase 10.21468/SciPostPhys.8.3.046} {\bibfield
  {journal} {\bibinfo  {journal} {SciPost Phys.}\ }\textbf {\bibinfo {volume}
  {8}},\ \bibinfo {pages} {046} (\bibinfo {year} {2020})}\BibitemShut {NoStop}%
\bibitem [{\citenamefont {Niezgoda}\ \emph {et~al.}(2020)\citenamefont
  {Niezgoda}, \citenamefont {Panfil},\ and\ \citenamefont
  {Chwede\ifmmode~\acute{n}\else \'{n}\fi{}czuk}}]{PhysRevA.102.042206}%
  \BibitemOpen
  \bibfield  {author} {\bibinfo {author} {\bibfnamefont {A.}~\bibnamefont
  {Niezgoda}}, \bibinfo {author} {\bibfnamefont {M.}~\bibnamefont {Panfil}}, \
  and\ \bibinfo {author} {\bibfnamefont {J.}~\bibnamefont
  {Chwede\ifmmode~\acute{n}\else \'{n}\fi{}czuk}},\ }\href {\doibase
  10.1103/PhysRevA.102.042206} {\bibfield  {journal} {\bibinfo  {journal}
  {Phys. Rev. A}\ }\textbf {\bibinfo {volume} {102}},\ \bibinfo {pages}
  {042206} (\bibinfo {year} {2020})}\BibitemShut {NoStop}%
\bibitem [{\citenamefont {Alet}\ \emph {et~al.}(2007)\citenamefont {Alet},
  \citenamefont {Capponi}, \citenamefont {Laflorencie},\ and\ \citenamefont
  {Mambrini}}]{Alet2007}%
  \BibitemOpen
  \bibfield  {author} {\bibinfo {author} {\bibfnamefont {F.}~\bibnamefont
  {Alet}}, \bibinfo {author} {\bibfnamefont {S.}~\bibnamefont {Capponi}},
  \bibinfo {author} {\bibfnamefont {N.}~\bibnamefont {Laflorencie}}, \ and\
  \bibinfo {author} {\bibfnamefont {M.}~\bibnamefont {Mambrini}},\ }\href
  {\doibase 10.1103/PhysRevLett.99.117204} {\bibfield  {journal} {\bibinfo
  {journal} {Phys. Rev. Lett.}\ }\textbf {\bibinfo {volume} {99}},\ \bibinfo
  {pages} {117204} (\bibinfo {year} {2007})}\BibitemShut {NoStop}%
\bibitem [{\citenamefont {Korepin}\ \emph {et~al.}(1993)\citenamefont
  {Korepin}, \citenamefont {Bogoliubov},\ and\ \citenamefont
  {Izergin}}]{korepin_bogoliubov_izergin_1993}%
  \BibitemOpen
  \bibfield  {author} {\bibinfo {author} {\bibfnamefont {V.~E.}\ \bibnamefont
  {Korepin}}, \bibinfo {author} {\bibfnamefont {N.~M.}\ \bibnamefont
  {Bogoliubov}}, \ and\ \bibinfo {author} {\bibfnamefont {A.~G.}\ \bibnamefont
  {Izergin}},\ }\href {\doibase 10.1017/CBO9780511628832} {\emph {\bibinfo
  {title} {Quantum Inverse Scattering Method and Correlation Functions}}},\
  Cambridge Monographs on Mathematical Physics\ (\bibinfo  {publisher}
  {Cambridge University Press},\ \bibinfo {year} {1993})\BibitemShut {NoStop}%
\bibitem [{\citenamefont {Villain}(1975)}]{Villain1975}%
  \BibitemOpen
  \bibfield  {author} {\bibinfo {author} {\bibfnamefont {J.}~\bibnamefont
  {Villain}},\ }\href {\doibase https://doi.org/10.1016/0378-4363(75)90101-1}
  {\bibfield  {journal} {\bibinfo  {journal} {Physica B+C}\ }\textbf {\bibinfo
  {volume} {79}},\ \bibinfo {pages} {1} (\bibinfo {year} {1975})}\BibitemShut
  {NoStop}%
\bibitem [{SM()}]{SM}%
  \BibitemOpen
  \href@noop {} {}\bibinfo {note} {See Supplemental Material at (...) for
  details.}\BibitemShut {Stop}%
\bibitem [{\citenamefont {Vidal}\ \emph
  {et~al.}(2003{\natexlab{b}})\citenamefont {Vidal}, \citenamefont {Latorre},
  \citenamefont {Rico},\ and\ \citenamefont {Kitaev}}]{Vidal2003}%
  \BibitemOpen
  \bibfield  {author} {\bibinfo {author} {\bibfnamefont {G.}~\bibnamefont
  {Vidal}}, \bibinfo {author} {\bibfnamefont {J.~I.}\ \bibnamefont {Latorre}},
  \bibinfo {author} {\bibfnamefont {E.}~\bibnamefont {Rico}}, \ and\ \bibinfo
  {author} {\bibfnamefont {A.}~\bibnamefont {Kitaev}},\ }\href {\doibase
  10.1103/PhysRevLett.90.227902} {\bibfield  {journal} {\bibinfo  {journal}
  {Phys. Rev. Lett.}\ }\textbf {\bibinfo {volume} {90}},\ \bibinfo {pages}
  {227902} (\bibinfo {year} {2003}{\natexlab{b}})}\BibitemShut {NoStop}%
\bibitem [{\citenamefont {Zhao}\ \emph {et~al.}(2025)\citenamefont {Zhao},
  \citenamefont {Yang}, \citenamefont {Henriques}, \citenamefont
  {Ferri-Cortés}, \citenamefont {Catarina}, \citenamefont {Pignedoli},
  \citenamefont {Ma}, \citenamefont {Feng}, \citenamefont {Ruffieux},
  \citenamefont {Fernández-Rossier},\ and\ \citenamefont {Fasel}}]{Zhao2025}%
  \BibitemOpen
  \bibfield  {author} {\bibinfo {author} {\bibfnamefont {C.}~\bibnamefont
  {Zhao}}, \bibinfo {author} {\bibfnamefont {L.}~\bibnamefont {Yang}}, \bibinfo
  {author} {\bibfnamefont {J.~C.~G.}\ \bibnamefont {Henriques}}, \bibinfo
  {author} {\bibfnamefont {M.}~\bibnamefont {Ferri-Cortés}}, \bibinfo {author}
  {\bibfnamefont {G.}~\bibnamefont {Catarina}}, \bibinfo {author}
  {\bibfnamefont {C.~A.}\ \bibnamefont {Pignedoli}}, \bibinfo {author}
  {\bibfnamefont {J.}~\bibnamefont {Ma}}, \bibinfo {author} {\bibfnamefont
  {X.}~\bibnamefont {Feng}}, \bibinfo {author} {\bibfnamefont {P.}~\bibnamefont
  {Ruffieux}}, \bibinfo {author} {\bibfnamefont {J.}~\bibnamefont
  {Fernández-Rossier}}, \ and\ \bibinfo {author} {\bibfnamefont
  {R.}~\bibnamefont {Fasel}},\ }\href {\doibase 10.1038/s41563-025-02166-1}
  {\bibfield  {journal} {\bibinfo  {journal} {Nature Materials}\ } (\bibinfo
  {year} {2025}),\ 10.1038/s41563-025-02166-1}\BibitemShut {NoStop}%
\bibitem [{\citenamefont {Keselman}\ \emph {et~al.}(2020)\citenamefont
  {Keselman}, \citenamefont {Balents},\ and\ \citenamefont
  {Starykh}}]{Keselman2020}%
  \BibitemOpen
  \bibfield  {author} {\bibinfo {author} {\bibfnamefont {A.}~\bibnamefont
  {Keselman}}, \bibinfo {author} {\bibfnamefont {L.}~\bibnamefont {Balents}}, \
  and\ \bibinfo {author} {\bibfnamefont {O.~A.}\ \bibnamefont {Starykh}},\
  }\href {\doibase 10.1103/PhysRevLett.125.187201} {\bibfield  {journal}
  {\bibinfo  {journal} {Phys. Rev. Lett.}\ }\textbf {\bibinfo {volume} {125}},\
  \bibinfo {pages} {187201} (\bibinfo {year} {2020})}\BibitemShut {NoStop}%
\bibitem [{\citenamefont {Beekman}\ \emph {et~al.}(2019)\citenamefont
  {Beekman}, \citenamefont {Rademaker},\ and\ \citenamefont {van
  Wezel}}]{Beekman2019}%
  \BibitemOpen
  \bibfield  {author} {\bibinfo {author} {\bibfnamefont {A.~J.}\ \bibnamefont
  {Beekman}}, \bibinfo {author} {\bibfnamefont {L.}~\bibnamefont {Rademaker}},
  \ and\ \bibinfo {author} {\bibfnamefont {J.}~\bibnamefont {van Wezel}},\
  }\href {\doibase 10.21468/SciPostPhysLectNotes.11} {\bibfield  {journal}
  {\bibinfo  {journal} {SciPost Phys. Lect. Notes}\ ,\ \bibinfo {pages} {11}}
  (\bibinfo {year} {2019})}\BibitemShut {NoStop}%
\bibitem [{\citenamefont {Zaanen}\ and\ \citenamefont
  {Ole\ifmmode~\acute{s}\else \'{s}\fi{}}(1988)}]{Zaanen1988}%
  \BibitemOpen
  \bibfield  {author} {\bibinfo {author} {\bibfnamefont {J.}~\bibnamefont
  {Zaanen}}\ and\ \bibinfo {author} {\bibfnamefont {A.~M.}\ \bibnamefont
  {Ole\ifmmode~\acute{s}\else \'{s}\fi{}}},\ }\href {\doibase
  10.1103/PhysRevB.37.9423} {\bibfield  {journal} {\bibinfo  {journal} {Phys.
  Rev. B}\ }\textbf {\bibinfo {volume} {37}},\ \bibinfo {pages} {9423}
  (\bibinfo {year} {1988})}\BibitemShut {NoStop}%
\bibitem [{\citenamefont {Wohlfeld}\ \emph {et~al.}(2010)\citenamefont
  {Wohlfeld}, \citenamefont {Ole\ifmmode~\acute{s}\else \'{s}\fi{}},\ and\
  \citenamefont {Sawatzky}}]{Wohlfeld2010}%
  \BibitemOpen
  \bibfield  {author} {\bibinfo {author} {\bibfnamefont {K.}~\bibnamefont
  {Wohlfeld}}, \bibinfo {author} {\bibfnamefont {A.~M.}\ \bibnamefont
  {Ole\ifmmode~\acute{s}\else \'{s}\fi{}}}, \ and\ \bibinfo {author}
  {\bibfnamefont {G.~A.}\ \bibnamefont {Sawatzky}},\ }\href {\doibase
  10.1103/PhysRevB.81.214522} {\bibfield  {journal} {\bibinfo  {journal} {Phys.
  Rev. B}\ }\textbf {\bibinfo {volume} {81}},\ \bibinfo {pages} {214522}
  (\bibinfo {year} {2010})}\BibitemShut {NoStop}%
\bibitem [{\citenamefont {Hagendorf}\ and\ \citenamefont
  {Li{\'e}nardy}(2017)}]{Hagendorf_2017}%
  \BibitemOpen
  \bibfield  {author} {\bibinfo {author} {\bibfnamefont {C.}~\bibnamefont
  {Hagendorf}}\ and\ \bibinfo {author} {\bibfnamefont {J.}~\bibnamefont
  {Li{\'e}nardy}},\ }\href {\doibase 10.1088/1751-8121/aa67ff} {\bibfield
  {journal} {\bibinfo  {journal} {Journal of Physics A: Mathematical and
  Theoretical}\ }\textbf {\bibinfo {volume} {50}},\ \bibinfo {pages} {185202}
  (\bibinfo {year} {2017})}\BibitemShut {NoStop}%
\bibitem [{\citenamefont {Caux}\ \emph {et~al.}(2008)\citenamefont {Caux},
  \citenamefont {Mossel},\ and\ \citenamefont {Castillo}}]{Caux_2008}%
  \BibitemOpen
  \bibfield  {author} {\bibinfo {author} {\bibfnamefont {J.-S.}\ \bibnamefont
  {Caux}}, \bibinfo {author} {\bibfnamefont {J.}~\bibnamefont {Mossel}}, \ and\
  \bibinfo {author} {\bibfnamefont {I.~P.}\ \bibnamefont {Castillo}},\ }\href
  {\doibase 10.1088/1742-5468/2008/08/P08006} {\bibfield  {journal} {\bibinfo
  {journal} {Journal of Statistical Mechanics: Theory and Experiment}\ }\textbf
  {\bibinfo {volume} {2008}},\ \bibinfo {pages} {P08006} (\bibinfo {year}
  {2008})}\BibitemShut {NoStop}%
\bibitem [{\citenamefont {{Franchini}}(2017)}]{2017LNP...940.....F}%
  \BibitemOpen
  \bibfield  {author} {\bibinfo {author} {\bibfnamefont {F.}~\bibnamefont
  {{Franchini}}},\ }\href {\doibase 10.1007/978-3-319-48487-7} {\emph {\bibinfo
  {title} {{An Introduction to Integrable Techniques for One-Dimensional
  Quantum Systems}}}},\ Vol.\ \bibinfo {volume} {940}\ (\bibinfo {year}
  {2017})\BibitemShut {NoStop}%
\bibitem [{\citenamefont {Bohrdt}\ \emph {et~al.}(2018)\citenamefont {Bohrdt},
  \citenamefont {Greif}, \citenamefont {Demler}, \citenamefont {Knap},\ and\
  \citenamefont {Grusdt}}]{Bohrdt2018}%
  \BibitemOpen
  \bibfield  {author} {\bibinfo {author} {\bibfnamefont {A.}~\bibnamefont
  {Bohrdt}}, \bibinfo {author} {\bibfnamefont {D.}~\bibnamefont {Greif}},
  \bibinfo {author} {\bibfnamefont {E.}~\bibnamefont {Demler}}, \bibinfo
  {author} {\bibfnamefont {M.}~\bibnamefont {Knap}}, \ and\ \bibinfo {author}
  {\bibfnamefont {F.}~\bibnamefont {Grusdt}},\ }\href {\doibase
  10.1103/PhysRevB.97.125117} {\bibfield  {journal} {\bibinfo  {journal} {Phys.
  Rev. B}\ }\textbf {\bibinfo {volume} {97}},\ \bibinfo {pages} {125117}
  (\bibinfo {year} {2018})}\BibitemShut {NoStop}%
\bibitem [{\citenamefont {Rutkevich}(2022)}]{PhysRevB.106.134405}%
  \BibitemOpen
  \bibfield  {author} {\bibinfo {author} {\bibfnamefont {S.~B.}\ \bibnamefont
  {Rutkevich}},\ }\href {\doibase 10.1103/PhysRevB.106.134405} {\bibfield
  {journal} {\bibinfo  {journal} {Phys. Rev. B}\ }\textbf {\bibinfo {volume}
  {106}},\ \bibinfo {pages} {134405} (\bibinfo {year} {2022})}\BibitemShut
  {NoStop}%
\bibitem [{\citenamefont {Shastry}\ and\ \citenamefont
  {Sutherland}(1981)}]{Shastry1981}%
  \BibitemOpen
  \bibfield  {author} {\bibinfo {author} {\bibfnamefont {B.~S.}\ \bibnamefont
  {Shastry}}\ and\ \bibinfo {author} {\bibfnamefont {B.}~\bibnamefont
  {Sutherland}},\ }\href {\doibase 10.1103/PhysRevLett.47.964} {\bibfield
  {journal} {\bibinfo  {journal} {Phys. Rev. Lett.}\ }\textbf {\bibinfo
  {volume} {47}},\ \bibinfo {pages} {964} (\bibinfo {year} {1981})}\BibitemShut
  {NoStop}%
\bibitem [{\citenamefont {Lavar{\'e}lo}\ and\ \citenamefont
  {Roux}(2014)}]{Lavarelo2014}%
  \BibitemOpen
  \bibfield  {author} {\bibinfo {author} {\bibfnamefont {A.}~\bibnamefont
  {Lavar{\'e}lo}}\ and\ \bibinfo {author} {\bibfnamefont {G.}~\bibnamefont
  {Roux}},\ }\href@noop {} {\bibfield  {journal} {\bibinfo  {journal} {The
  European Physical Journal B}\ }\textbf {\bibinfo {volume} {87}},\ \bibinfo
  {pages} {229} (\bibinfo {year} {2014})}\BibitemShut {NoStop}%
\bibitem [{Note1()}]{Note1}%
  \BibitemOpen
  \bibinfo {note} {Here we apply the standard superexchange formula, [{\protect
  \it cf}.~\cite {Zaanen1988} or Eq. (7) in \cite {Wohlfeld2010}] to the bond
  with three transition metal ions.}\BibitemShut {Stop}%
\bibitem [{\citenamefont {Ament}\ \emph {et~al.}(2011)\citenamefont {Ament},
  \citenamefont {van Veenendaal}, \citenamefont {Devereaux}, \citenamefont
  {Hill},\ and\ \citenamefont {van~den Brink}}]{Ament2011}%
  \BibitemOpen
  \bibfield  {author} {\bibinfo {author} {\bibfnamefont {L.~J.~P.}\
  \bibnamefont {Ament}}, \bibinfo {author} {\bibfnamefont {M.}~\bibnamefont
  {van Veenendaal}}, \bibinfo {author} {\bibfnamefont {T.~P.}\ \bibnamefont
  {Devereaux}}, \bibinfo {author} {\bibfnamefont {J.~P.}\ \bibnamefont {Hill}},
  \ and\ \bibinfo {author} {\bibfnamefont {J.}~\bibnamefont {van~den Brink}},\
  }\href {\doibase 10.1103/RevModPhys.83.705} {\bibfield  {journal} {\bibinfo
  {journal} {Rev. Mod. Phys.}\ }\textbf {\bibinfo {volume} {83}},\ \bibinfo
  {pages} {705} (\bibinfo {year} {2011})}\BibitemShut {NoStop}%
\bibitem [{\citenamefont {Jia}\ \emph {et~al.}(2016)\citenamefont {Jia},
  \citenamefont {Wohlfeld}, \citenamefont {Wang}, \citenamefont {Moritz},\ and\
  \citenamefont {Devereaux}}]{Jia2016}%
  \BibitemOpen
  \bibfield  {author} {\bibinfo {author} {\bibfnamefont {C.}~\bibnamefont
  {Jia}}, \bibinfo {author} {\bibfnamefont {K.}~\bibnamefont {Wohlfeld}},
  \bibinfo {author} {\bibfnamefont {Y.}~\bibnamefont {Wang}}, \bibinfo {author}
  {\bibfnamefont {B.}~\bibnamefont {Moritz}}, \ and\ \bibinfo {author}
  {\bibfnamefont {T.~P.}\ \bibnamefont {Devereaux}},\ }\href {\doibase
  10.1103/PhysRevX.6.021020} {\bibfield  {journal} {\bibinfo  {journal} {Phys.
  Rev. X}\ }\textbf {\bibinfo {volume} {6}},\ \bibinfo {pages} {021020}
  (\bibinfo {year} {2016})}\BibitemShut {NoStop}%
\bibitem [{\citenamefont {Sawatzky}\ \emph {et~al.}(2009)\citenamefont
  {Sawatzky}, \citenamefont {Elfimov}, \citenamefont {van~den Brink},\ and\
  \citenamefont {Zaanen}}]{Sawatzky2009}%
  \BibitemOpen
  \bibfield  {author} {\bibinfo {author} {\bibfnamefont {G.~A.}\ \bibnamefont
  {Sawatzky}}, \bibinfo {author} {\bibfnamefont {I.~S.}\ \bibnamefont
  {Elfimov}}, \bibinfo {author} {\bibfnamefont {J.}~\bibnamefont {van~den
  Brink}}, \ and\ \bibinfo {author} {\bibfnamefont {J.}~\bibnamefont
  {Zaanen}},\ }\href {\doibase 10.1209/0295-5075/86/17006} {\bibfield
  {journal} {\bibinfo  {journal} {EPL (Europhysics Letters)}\ }\textbf
  {\bibinfo {volume} {86}},\ \bibinfo {pages} {17006} (\bibinfo {year}
  {2009})}\BibitemShut {NoStop}%
\end{thebibliography}
\end{document}